\documentclass[aps,prx,a4paper,reprint,twocolumn,floatfix,superscriptaddress,notitlepage,usenames,dvipsnames,svgnames,x11names,table,nofootinbib,longbibliography]{revtex4-2}
\pdfoutput=1
\usepackage[T1]{fontenc}
\usepackage{graphicx}
\usepackage{amssymb,amsmath}

\usepackage{braket}
\usepackage{placeins}
\usepackage{tabularx}
\usepackage{siunitx}
\usepackage{physics}
\usepackage{makecell}
\usepackage{appendix}
\usepackage{soul}
\AtBeginDocument{\RenewCommandCopy\qty\SI}

\usepackage{pifont}
\usepackage{mathtools}

\makeatletter
\newcommand*{\addFileDependency}[1]{
\typeout{(#1)}

\@addtofilelist{#1}

\IfFileExists{#1}{}{\typeout{No file #1.}}
}\makeatother

\newcommand{\Sb}{^{123}\rm Sb}

\usepackage{xr}
\usepackage{soul}

\usepackage[pdfusetitle]{hyperref}
\hypersetup{pdflang={English},colorlinks=true,linkcolor=Cerulean,citecolor=Cerulean,urlcolor=Cerulean,breaklinks=true}

\usepackage[nameinlink,capitalize,capitalise]{cleveref}

\begin{document}

\title{Dynamical Decoupling of Nonclassical States in a Biased-Noise Spin Qudit}

\author{Arjen Vaartjes}
\thanks{These authors contributed equally}
\affiliation{School of Electrical Engineering and Telecommunications, UNSW Sydney, Sydney, NSW 2052, Australia}

\author{Namit Anand}
\thanks{These authors contributed equally}

\affiliation{HPE Quantum, Emergent Machine Intelligence, Hewlett Packard Labs, CA, USA}
\affiliation{Quantum Artificial Intelligence Laboratory (QuAIL), NASA Ames Research Center, Moffett Field, CA, 94035, USA}
\affiliation{KBR, Inc., 601 Jefferson St., Houston, TX 77002, USA}

\author{Rocky Y. Su}
\affiliation{School of Electrical Engineering and Telecommunications, UNSW Sydney, Sydney, NSW 2052, Australia}

\author{Akram Youssry}
\affiliation{School of Electrical Engineering and Telecommunications, UNSW Sydney, Sydney, NSW 2052, Australia}
\affiliation{ARC Centre of Excellence for Quantum Computer Performance \& Integration (QPI)}

\author{Ritik Sareen}
\affiliation{School of Electrical Engineering and Telecommunications, UNSW Sydney, Sydney, NSW 2052, Australia}
\affiliation{ARC Centre of Excellence for Quantum Computer Performance \& Integration (QPI)}
\affiliation{School of Engineering, RMIT University, Melbourne, VIC, 3000, Australia}

\author{Jeffrey Marshall}

\affiliation{Quantum Artificial Intelligence Laboratory (QuAIL), NASA Ames Research Center, Moffett Field, CA, 94035, USA}
\affiliation{USRA Research Institute for Advanced Computer Science, Mountain View, California 94043, USA}

\author{Mark R. van Blankenstein}
\affiliation{School of Electrical Engineering and Telecommunications, UNSW Sydney, Sydney, NSW 2052, Australia}
\affiliation{ARC Centre of Excellence for Quantum Computer Performance \& Integration (QPI)}

\author{Danielle Holmes}
\affiliation{School of Electrical Engineering and Telecommunications, UNSW Sydney, Sydney, NSW 2052, Australia}
\affiliation{ARC Centre of Excellence for Quantum Computer Performance \& Integration (QPI)}

\author{Benjamin Wilhelm}
\affiliation{School of Electrical Engineering and Telecommunications, UNSW Sydney, Sydney, NSW 2052, Australia}
\affiliation{ARC Centre of Excellence for Quantum Computer Performance \& Integration (QPI)}

\author{Bibek Pokharel}

\affiliation{Department of Physics \& Astronomy \& Center for Quantum Information Science \& Technology, University of Southern California, Los Angeles, CA 90089, USA}

\author{Xi Yu}
\affiliation{School of Electrical Engineering and Telecommunications, UNSW Sydney, Sydney, NSW 2052, Australia}

\author{Laura A. O'Neill}
\affiliation{School of Electrical Engineering and Telecommunications, UNSW Sydney, Sydney, NSW 2052, Australia}
\affiliation{ARC Centre of Excellence for Quantum Computer Performance \& Integration (QPI)}

\author{Martin Nurizzo}
\affiliation{School of Electrical Engineering and Telecommunications, UNSW Sydney, Sydney, NSW 2052, Australia}

\author{Alexander M. Jakob}
\affiliation{School of Physics, University of Melbourne, Melbourne, VIC 3010, Australia}

\author{Fay E. Hudson}
\affiliation{School of Electrical Engineering and Telecommunications, UNSW Sydney, Sydney, NSW 2052, Australia}
\affiliation{Diraq Pty. Ltd., Sydney, NSW, Australia}

\author{Kohei M. Itoh}
\affiliation{School of Fundamental Science and Technology, Keio University, Kohoku-ku, Yokohama, Japan}

\author{Andrew S. Dzurak}
\affiliation{School of Electrical Engineering and Telecommunications, UNSW Sydney, Sydney, NSW 2052, Australia}
\affiliation{Diraq Pty. Ltd., Sydney, NSW, Australia}

\author{David N. Jamieson}
\affiliation{School of Physics, University of Melbourne, Melbourne, VIC 3010, Australia}

\author{Andrea Morello}
\affiliation{School of Electrical Engineering and Telecommunications, UNSW Sydney, Sydney, NSW 2052, Australia}
\affiliation{ARC Centre of Excellence for Quantum Computer Performance \& Integration (QPI)}

\date{\today}

\begin{abstract}
Encoding quantum information in $d$-dimensional systems (qudits) with $d>2$ offers appealing new directions for resource-efficient quantum computing. However, protecting the coherence of nonclassical qudit resource states remains an open challenge. Dynamical decoupling (DD) is a well-established technique for extending coherence, but its application to qudits is hindered by the quadratic scaling of the decoupling sequence overhead with the dimension $d$. As a result, demonstrations of qudit DD have so far been limited to low-dimensional qudits. Here, we show that noise-biased qudits enable a nontrivial reduction of this overhead, for example, from quadratic to linear. We perform qudit DD on a model $d=8$ qudit, the nuclear spin of an antimony donor in silicon. We benchmark DD performance on highly nonclassical states such as uniform superpositions, cat states, and Haar random states and measure coherence improvements up to a factor $5$ to $10$. For a spin-cat error correction code, continuous DD extends the coherence time of the logical superposition $\ket{+}_L$ to $\SI{225}{\milli \second}$, compared to $\SI{19}{\milli \second}$ under free evolution and $\SI{85}{\milli \second}$ with a Hahn echo. This improvement shows that DD sequences tailored to a system's noise bias can significantly enhance the lifetime of encoded logical qubits.
\end{abstract}

\maketitle


\section{Introduction}
Dynamical decoupling (DD)~\cite{Viola1998,viola1999dynamical,zanardi1999symmetrizing,duan1999suppressing} can suppress errors in both pre-fault-tolerant and fault-tolerant quantum computation~\cite{ng2011combining, paz2013optimally}. It originates from the field of Nuclear Magnetic Resonance (NMR) \cite{hahn1950spin,meiboom1958modified,haeberlen}, and has grown to become an essential tool in quantum computing and quantum sensing \cite{suter2016colloquium}. It has been extensively employed to extend the coherence times in several qubit platforms, including trapped ions~\cite{biercuk2009optimized}, nitrogen-vacancy (NV) centers in diamond~\cite{DeLange2010}, superconducting qubits~\cite{cywinski2008enhance,bylander2011noise,ezzell2023dynamical}, spins in semiconductors~\cite{medford2012scaling,Muhonen2014}, and neutral atoms~\cite{Chiu2025ContinuousSystem}.
For single qubits, decoupling sequences with optimal performance under various noise profiles and pulse imperfections have been extensively analyzed~\cite{viola2003robust,biercuk2009optimized,du2009preserving,Uhrig2011,Ajoy2011,souza2011robust,genov2017arbitrarily,ezzell2023dynamical,d2026proof}.
Recent works on qubit DD therefore focus on new applications such as rapid qubit characterization \cite{Tripathi2025characterization}, scaling to larger systems by mitigating crosstalk \cite{Tripathi2022,tong2025empirical,tong2026learning,coote2025resource} and incorporating DD in error correction schemes \cite{paz2013optimally,Han2025,kasatkin2026quantum}. 

In contrast with standard, qubit-based quantum computing, high-dimensional quantum computing \cite{wang2020qudits} seeks to exploit physical systems with a Hilbert space dimension $d>2$ (`qudits'). Qudit-based quantum computing offers appealing advantages, such as reducing hardware overheads and simplifying circuit compilation. However, it often requires a drastic rethinking of how quantum operations are performed. DD is a prime example of how qudits behave differently from qubits.

Here we present a comprehensive study of the application of DD techniques to a model qudit system. Qudit DD differs from qubit-based methods due to the presence of higher-order noise operators and the larger size of their decoupling groups. 

Whereas qubit DD strategies rely on the well-known Pauli group, for qudits the corresponding universal decoupling group is the Heisenberg-Weyl (HW) group (see Supplementary Information S2.C), whose size scales as $d^3$ with the qudit dimension $d$~\cite{lidar_lecture_2020}.


Crucially, the number of operations for decoupling an arbitrary qudit system-bath Hamiltonian, i.e. \emph{universal} DD, scales as \(d^2\), not \(d^3\), since the \(d\) phase factors in the HW group are physically irrelevant for DD. Nonetheless, the \(d^2\) scaling makes universal DD sequences experimentally demanding for large $d$. Qudit DD has been implemented in superconducting qubits~\cite{Tripathi2025QuditProcessor, iiyama2024qudit}, trapped ions~\cite{yuan2022preserving} and NV centers~\cite{zhou2024robust}, but the dimension of the qudit has so far been limited to $d=4$. Increasing the dimension to $d=8$ would make a universal DD sequence consist of 64 unitaries, whereas the universal qubit DD sequence (XYXY) requires only 4~\cite{Gullion1990}.

In this work, we study a $\Sb$ nuclear spin qudit ($d=8$) in silicon, which displays an extreme noise bias: the system is many orders of magnitude more sensitive to dephasing ($T_2$-errors) compared to relaxation ($T_1$-errors)~\cite{Savytskyy2023,yu2025schrodinger}. This inherent noise asymmetry has two attractive consequences. 
First, it makes high-spin nuclei a suitable platform for encoding biased-noise logical cat qubits \cite{Gross2024,yu2025schrodinger}.
Second, the same noise bias greatly simplifies the design of DD sequences. Instead of cycling over the entire Heisenberg-Weyl group, DD only needs to suppress a few noise channels when energy relaxation is negligible. 
Here we analyze a cyclic sequence that can protect against arbitrary order diagonal noise. 

Our approach is to cyclically permute all 8 spin eigenstates using shift unitaries ($X_8$)~\cite{Tripathi2025QuditProcessor}, canceling all noise that couples to diagonal dephasing operators. We experimentally implement the \textit{cyclic-DD} sequence and validate it for various nonclassical states such as uniform superpositions and Schr\"odinger cat states, achieving a factor 5-10 increase in the coherence time. Moreover, we extend this validation to Haar-random states, demonstrating that the cyclic-DD sequence provides protection across states deep in the hard-to-reach regions of the Hilbert space. Our work introduces the practice of assessing DD performance on Haar-random states as a novel and rigorous benchmarking procedure to certify the true performance of DD sequences, which cannot be explored by benchmarking on simple initial states alone.

We then go beyond protecting arbitrary qudit states and ask how DD can be tailored to protect quantum information encoded in the qudit. In particular, we consider the spin-cat code, for which the logical states occupy a structured subspace within the $d=8$ Hilbert space. This structure enables DD strategies that are adapted to the logical states rather than the full qudit. We compare four strategies: \textit{Hahn Echo, Cyclic DD, Rotary Echo and Spin Locking}, the latter two being continuously driven DD techniques. We find that \textit{spin locking}, implemented by driving continuously around a common symmetry axis of the logical states, performs best and provides more than an order-of-magnitude improvement of the lifetimes of the logical code words. These results show the benefit of applying a DD strategy that is tailored to the specific structure of the logic states.

\section{Qudit DD}
\subsection{Cyclic permutations}

We use dimensionless spin operators and Hamiltonians in angular-frequency units, so $U(t)=e^{-i H t}$. Consider a single qudit system with a system-bath coupling of the form,
\begin{equation}
H_{sb} = \sum\limits_{j=1}^{d^2-1} \sigma_{j} \otimes b_j
\end{equation}
where \(\{ \sigma_{j} \}\) form a a traceless Hermitian orthonormal basis for the qudit operator space (along with the identity operator) and \(\{ b_{j} \}\) are operators acting on the bath. The key idea behind DD is to apply a sequence of pulses on the qudit such that the system-bath coupling is time-averaged to $H_{s b} \mapsto \mathbb{I}_d \otimes \operatorname{Tr}_s\left[H_{s b}\right] / d$, at leading Magnus order. Namely, the effective system Hamiltonian is averaged to a scalar. As a result, one has ``dynamically decoupled'' the system and the bath, hence the apropos name~\cite{lidar2014review}. If the system-bath coupling has a simple form (e.g., with a strong noise-bias) then one can find a minimal decoupling group to achieve this cancellation, e.g., using Hahn echo for pure dephasing, as we show later.

DD pulse sequences cycle over $N$ unitaries in the decoupling group $\mathcal{G}$, resulting in an effective evolution of the form,
\begin{equation}
    U(t) = \prod_{i=0}^{N}g_i^{\dagger}f_t g_i \hspace{1cm} 
\end{equation}
where $g_i\in\mathcal{G}$ and \(f_t = e^{-i H_{sb} t}\). For single qubits, the universal DD group is formed by the Pauli group $\mathcal{G}=\{I, \sigma_x, \sigma_y, \sigma_z\}$. More generally, extending to the qudit case, the universal decoupling group is formed by a unitary 1-design such as the Heisenberg-Weyl group~\cite{Tripathi2025QuditProcessor,anand2026quditdesigns}.

If the noise is dominated by dephasing, i.e. $f_\tau \propto e^{-i\sigma_z \tau}$, cycling over the subgroup $\{I, \sigma_x\}$ of the Pauli group is sufficient:
\begin{equation}
    \sigma_x-f_\tau-\sigma_x-f_\tau.
    \label{eq:hahn_echo}
\end{equation}

This is the well-known Hahn echo DD sequence \cite{hahn1950spin}, which relies on the anti-commutation of the system-bath operators and the pulse unitaries: $\{\sigma_x, \sigma_z\}=0$, leading to an inversion of the system-bath interaction (see Supplementary Information S2.A). A useful insight is that sequence~(\ref{eq:hahn_echo}) can be viewed as a cyclic permutation over the two qubit eigenstates, averaging the diagonal $\sigma_z$ operator.

Intuitively, the Hahn echo can be generalized to qudits using global $SU(2)$-rotations~\cite{Yu2024, Champion2024, Roy2024} (see Supplementary Information S2.B), which cancel linear dephasing terms proportional to $I_z$. However, in high-dimensional systems, the system-bath interaction generally includes higher-order diagonal operators such as $I_z^2$, which are left unchanged by the $SU(2)$ Hahn echo (see Supplementary Information S2.B). 

To suppress arbitrary diagonal noise, it suffices to average over the cyclic group of permutations $\mathcal{G_{\rm cyclic}}=\{X_d^k: k=0, ...,d-1\}$~\cite{zanardi1999symmetrizing, sun2024permutational}. The generator of this group is the Shift operator $X_d$ (see Supplementary Information S2.C), which for $d=8$ takes the form:
\begin{equation}
    X_8 = 
    \begin{pmatrix}
        0 & 1 & 0 & 0 & 0 & 0 & 0 & 0 \\
        0 & 0 & 1 & 0 & 0 & 0 & 0 & 0 \\
        0 & 0 & 0 & 1 & 0 & 0 & 0 & 0 \\
        0 & 0 & 0 & 0 & 1 & 0 & 0 & 0 \\
        0 & 0 & 0 & 0 & 0 & 1 & 0 & 0 \\
        0 & 0 & 0 & 0 & 0 & 0 & 1 & 0 \\
        0 & 0 & 0 & 0 & 0 & 0 & 0 & 1 \\
        1 & 0 & 0 & 0 & 0 & 0 & 0 & 0 \\
    \end{pmatrix}
\end{equation}
Cycling over this group (i.e. applying this unitary $d=8$ times) makes each diagonal entry visit all $d$ positions exactly once, so for any diagonal operator $D$:
\begin{equation}
    \frac{1}{d}\sum_{k=0}^{d-1}(X_d^k)^\dagger D X^k_d = \frac{{\rm Tr}(D)}{d} \mathbb{I}_d.
\end{equation}

The reduction of the decoupling sequence overhead arises from the noise bias, which manifests as the dominant system-bath interactions being limited to diagonal (system) operators. Arbitrary noise generally requires $d^2$ decoupling unitaries, whereas for dephasing noise, the cyclic group suffices with only $d$ unitaries. In the lab, each $X_d$ unitary can be constructed with $d-1$ native pulses, as we discuss in Fig.~\ref{fig:2}, reducing the total physical pulse overhead from $d^2(d-1)$ to $d(d-1)$ (see Supplementary Information S2.F for a detailed discussion).

Further reduction of the decoupling sequences can be achieved if the quantum states that need to be protected are limited to subspaces of $d$-dimensional Hilbert space. If the subspace only spans $k$ eigenstates, only those states involved need to be permuted, provided that noise and control preserve this subspace. In Section IV of this paper we utilize this to achieve a drastic increase in coherence times in an encoded cat qubit with relatively simple decoupling sequences.


\section{$\Sb$ Noise Model}

\begin{figure}
    \centering
    \includegraphics[width=\linewidth]{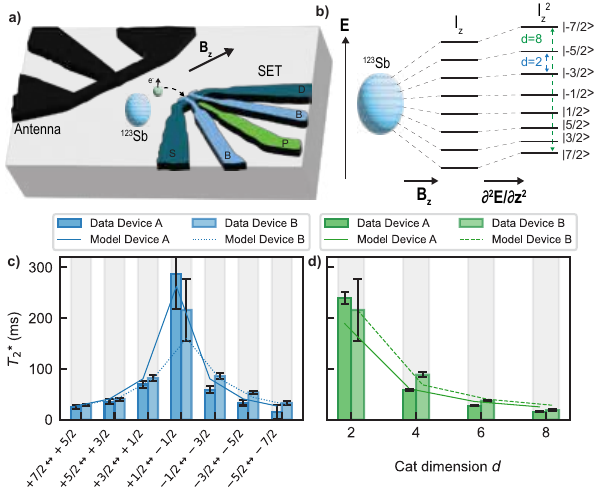}
    \caption{\textbf{Noise in the $\Sb$ nuclear qudit.} 
    \textbf{a)} Device schematic. A single $\Sb$ implanted in a silicon chip binds a hyperfine-coupled electron, tunnel-coupled to a readout Single Electron Transistor (SET), formed by a source (S)-drain (D) channel, two barrier (B) gates and a plunger (P) gate. An on-chip antenna provides an oscillating magnetic field to drive both Electron Spin Resonance (ESR) and Nuclear Magnetic Resonance (NMR). An externally applied magnetic field $B_z$ defines the spin quantization axis. 
    \textbf{b)} Energy diagram of the spin states of the $\Sb$ nucleus. The magnetic field $B_z$ coupled to the linear $I_z$ operator splits the 8 eigenstates equally. An electric field gradient $\partial^2 E/\partial \alpha \partial \beta$ couples to the quadratic $I_z^2$ operator through the quadrupole interaction and parabolically separates the eigenenergies.
    \textbf{c)} Measured and simulated Ramsey coherence times $T_2^*$ for all 7 adjacent qubit subspaces within the qudit for two devices (A and B). The outer subspaces show significantly shorter $T_2^*$ compared to middle subspaces due to increased sensitivity to electric $I_z^2$ noise. 
    \textbf{d)} Measured and simulated Ramsey experiments for symmetric $d$-dimensional cat states show a decrease in $T_2^*$ due to increased magnetic noise sensitivity.}
    \label{fig:1}
\end{figure}

In this work, we perform DD experiments on a model $d=8$ qudit, the spin-7/2 $\Sb$ nucleus \cite{Asaad2020,Fernandez2024,yu2025schrodinger} in silicon \cite{Itoh2014} (Fig.~\ref{fig:1}a).
We apply an external magnetic field $\mathbf{B_0}=1.395~\si{\tesla}~\mathbf{\hat{z}}$ in plane with the chip to Zeeman-split the nuclear energy levels (Fig.~\ref{fig:1}b). A second-order energy splitting, which serves a similar function as the anharmonicity in superconducting transmon qubits and qudits \cite{Champion2024}, is provided by the quadrupole interaction which couples electric field gradients (EFG) to the quadratic spin operators \cite{Asaad2020} (see Equation~\ref{eq:system_hamiltonian}). The quadrupole splitting of $\Delta f_q=\SI{-18.0 \pm 0.9}{\kilo \hertz}$~\cite{vaartjes2026maximizing} makes transitions between each pair of qudit levels individually addressable. 

Noise enters the system through fluctuations of the electric and magnetic fields. These noise channels are captured by a Hamiltonian of the form:

\begin{equation}
    H = H_s + H_b + H_{sb},
\end{equation}

where $H_{\rm s}$ is the system ($\Sb$ nucleus) Hamiltonian, $H_{\rm b}$ describes the noise bath, and $H_{\rm sb}$ is the system-bath interaction. The system Hamiltonian is given by \cite{Asaad2020}:

\begin{equation}  
H_{\rm s} = -\gamma_n B_z I_z + \sum_{\alpha, \beta \in \{x,y,z\}} Q_{\alpha\beta} I_\alpha I_\beta
\label{eq:system_hamiltonian}
\end{equation}

This Hamiltonian shows that fluctuations in the magnetic field $B_z$ couple through the $I_z$ operator. The second term in the Hamiltonian describes the quadrupolar interaction. The quadrupole tensor $Q$ depends on the electric field gradient $V$:

\begin{equation}
    Q_{\alpha\beta} = \frac{e Q}{2I(2I-1)\hbar} V_{\alpha\beta},
\end{equation}
where $V_{\alpha\beta}$ is the second partial derivative of the electric field: $V_{\alpha\beta} = \partial^2 E/\partial \alpha \partial\beta$. Electric field fluctuations are converted to electric field gradients via the gradient-electric tensor $R$ \cite{Asaad2020}, and propagate as fluctuations in the second-order terms $I_\alpha I_\beta$.
In the presence of a strong external magnetic field, the static Zeeman interaction ($\gamma_n B_z/(2\pi) \approx$ \SI{8}{\mega \hertz}) dominates over the quadrupole coupling ($\sim$ $\SI{20}{\kilo \hertz}$)~\cite{vaartjes2026maximizing}. We therefore apply the secular approximation~\cite{Chizhik2014}, leaving only the terms that commute with $I_z$, i.e. we treat the system-bath interaction as pure dephasing with the following time-dependent Hamiltonian:

\begin{equation}
    H_{\rm sb} = \beta_z(t)I_z \otimes b_1+ \beta_{zz}(t)I_z^2 \otimes 
    b_2,
\end{equation}
where $\beta_z(t)$ and $\beta_{zz}(t)$ are time-dependent coupling rates of magnetic and electric noise and $b_{1,2}$ are (unknown) bath operators. For simulations, we assume the bath couplings $\beta_z$ and $\beta_{zz}$ are independent zero-mean Gaussian processes with power spectra specified in Supplementary Information S5.

\subsection{Separating magnetic ($I_z$) and electric ($I_z^2$) noise}
\label{sec:characterizing_noise}

We performed Ramsey experiments to assess the relative contributions of electric an magnetic noise. The measurements were repeated on two nominally identical devices, referred to as Device A and Device B. We then fit the data with a noise model based on cumulant expansion, treated extensively in Supplementary S5.  

\subsubsection{Characterizing electric $I_z^2$ noise with qubit subspaces}

Figure~\ref{fig:1}c shows dephasing times of qubits defined within $d=2$ subspaces of the $\Sb$ qudit,  e.g. $\{\ket{-7/2}, \ket{-5/2}\}$. We observed a clear trend whereby $T_2^*$ decreases by moving away from the central transition $\{\ket{-1/2}\leftrightarrow \ket{1/2}\}$. This is the signature of the impact of electric noise, which enters the system through the quadratic $I_z^2$ term. 

We explain this effect by analyzing the differences in expectation values of $I_z$ and $I_z^2$ for each qubit subspace (Tab.~\ref{tab:expectation_values_noise}). The difference $|\Delta I_z|=1$ between $I_z$ eigenvalues of adjacent spin states is the same for all qubit subspaces. If the noise in the $\Sb$ system were purely magnetic, all qubit subspaces should be affected equally and we would expect the same $T_2^*$ for all subspaces. 

\begin{table}[h!]
\centering
\begin{tabular}{|c|c|c|}
\hline
NMR Transition & $|\Delta\langle I_z \rangle|$ & $|\Delta\langle I_z^2 \rangle|$ \\ \hline
$-7/2 \leftrightarrow -5/2$ & 1 & $49/4 - 25/4 = 6$ \\
$-5/2 \leftrightarrow -3/2$ & 1 & $25/4 - 9/4 = 4$ \\
$-3/2 \leftrightarrow -1/2$ & 1 & $9/4 - 1/4 = 2$ \\
$-1/2 \leftrightarrow 1/2$  & 1 & $1/4 - 1/4 = 0$ \\
$1/2 \leftrightarrow 3/2$   & 1 & $9/4 - 1/4 = 2$ \\
$3/2 \leftrightarrow 5/2$   & 1 & $25/4 - 9/4 = 4$ \\
$5/2 \leftrightarrow 7/2$   & 1 & $49/4 - 25/4 = 6$ \\
\hline
\end{tabular}
\label{tab:expectation_values_noise}
\caption{Absolute changes in expectation values of \( I_z \) and \( I_z^2 \) for NMR transitions in a spin-7/2 system.}
\end{table}

Moving away from the center transition yields an increasing difference in the $I_z^2$ expectation values, which corresponds to an increased sensitivity to electric noise for the outer subspaces.
Notably, the eigenvalues of $I_z^2$ for the $\ket{-1/2} \leftrightarrow \ket{1/2}$ qubit subspace are equal for both states. As a result, the $\mathrm{span} \{\ket{-1/2},\ket{1/2}\}$ is a decoherence-free subspace for electric noise propagating through the $I_z^2$ operator. This transition is thus only sensitive to magnetic noise, and and shows a much longer $T_2^*$ compared to all others. 

\subsubsection{Quantifying the magnetic $I_z$ noise using cat states}

To isolate the magnetic noise of the $\Sb$ system, we perform Ramsey experiments on $d$-dimensional cat states. These states are symmetric superpositions of extremal spin eigenstates of the Hamiltonian, e.g. $\frac{1}{\sqrt{2}}(\ket{-5/2} +\ket{5/2})$ for $d=6$. Each even-$d$ cat state for different values of \(m\) lies in a separate decoherence-free subspace for electric noise, $\mathrm{span} \{|m\rangle, |-m\rangle\}$, being an eigenstate of $I_z^2$. Their dephasing arises solely from fluctuations in the magnetic field through the $I_z$ operator. 
The measurements in Fig.~\ref{fig:1}d show a decrease in $T_2^*$ as a function of $d$ between the eigenstates. This is consistent with the increased effect of magnetic noise on the energies of the $\ket{\pm m}$ eigenstates involved in the superposition, as $d=2m+1$ increases. 

We can reproduce all measured $T_2^*$ data (see fit lines in Fig.~\ref{fig:1}c-d) with the cumulant expansion noise model described in the Supplementary Information, section S5.

\begin{figure}
    \centering
    \includegraphics[width=\linewidth]{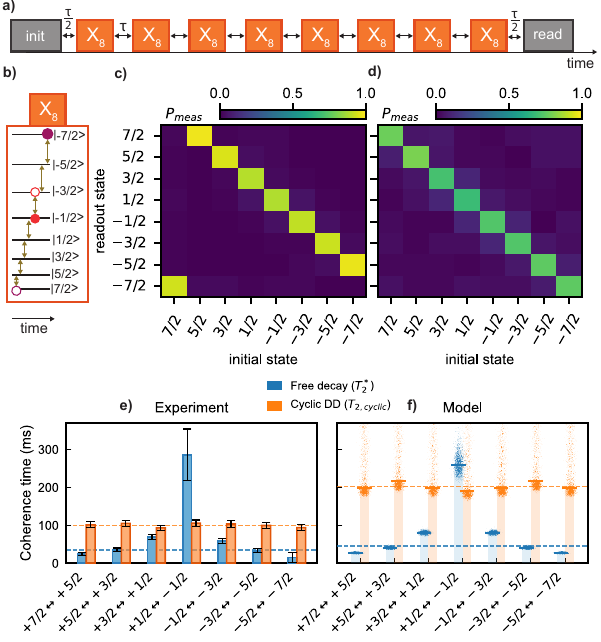}
    \caption{\textbf{Cyclic-DD sequence and performance on qubit superpositions.}
    \textbf{a)} Cyclic-DD pulse sequence, consisting of 8 $X_8$ operations cyclically permuting over all eigenstates. Each operation is separated by a free evolution time $\tau$ with initial and final segments of $\tau/2$ to form a symmetric sequence.
    \textbf{b)} One $d=8$ shift operation is constructed ($X_8$) by sequentially applying seven $\pi$ pulses from bottom to top. As a result, $\ket{7/2}$ population (purple) transfers all the way to $\ket{-7/2}$ whereas all other eigenstates move by $\Delta m = +1$, i.e. $\ket{-3/2}\rightarrow\ket{-1/2}$ (orange). 
    \textbf{c)} Measured transfer matrix of one $X_8$ operation.
    \textbf{d)} Measured transfer matrix of 8 sequential $X_8$ operations. Applying 8 $X_8$'s ideally reduces to the identity matrix. 
    \textbf{e)} Experimental results of the cyclic DD sequence on qubit subspaces compared to free decay. We observe that the DD sequence equalizes all $T_2$'s and decreases the mean decay rate $\langle\Gamma_2\rangle=\langle1/T_2\rangle$ (dashed horizontal lines show $1/\langle\Gamma_2\rangle$).
    \textbf{f)} Modeled coherence times for the free decay and cyclic DD sequences. The swarms display the distribution of 2000 simulation runs, and the horizontal bars indicate the median. }
    
    \label{fig:2}
\end{figure}

\section{Results}

\subsection{Cyclic DD on qubit subspaces}

We begin by applying the cyclic DD sequence (Fig.~\ref{fig:2}a) on the superpositions of adjacent eigenstates, i.e. qubit subspaces, ($\Delta m=1$) superpositions on Device A. The sequence effectively equalizes the coherence time of different qubit subspaces because it traverses all $d=2$ superposition states. 
This principle is illustrated in Fig.\ref{fig:2}b-c, which displays the pulse sequence and corresponding transfer matrix of one $d=8$ shift ($X_8$) operation. 
A single $X_8$ unitary is constructed by seven two-level $\pi$-pulses applied in an ascending ladder sequence, starting from $\ket{7/2}\leftrightarrow\ket{5/2}$. 
Under the effect of this pulse sequence, a state initially in $\ket{7/2}$ climbs the entire spin ladder to $\ket{-7/2}$, whereas any other state moves by $\Delta m = +1$, e.g. $\ket{-3/2}\rightarrow\ket{-1/2}$. We verify this behavior in Fig.~\ref{fig:2}c, which shows the measured transfer matrix of a single $X_8$. We confirm the cyclic nature of this permutation by applying it 8 times, and verify that we indeed retrieve the initial populations (Fig.~\ref{fig:2}d). 
In Fig.~\ref{fig:2}e, we display the result of the cyclic DD experiment on superpositions states in qubit subspaces. We observe an equalization of $T_2$, and measure a factor 2 improvement over the average of the dephasing rates $\Gamma_2=1/T_2^*$ of the individual qubit subspaces. The simulation with the cumulant expansion noise model (Fig.~\ref{fig:2}f) shows qualitatively similar behavior, although it predicts quantitatively longer coherence times than observed in the experiment. This is likely due to the fact that the total pulse duration of $12.8~\si{\milli\second}$ is relatively long compared to the total wait time. The finite pulse duration is not taken into account in the model, which assumes instantaneous pulses.

We attribute the limited $T_2$ improvement upon applying the cyclic DD sequence to two factors.
First, due to the cyclic nature of the DD sequence, any initial superposition state eventually reaches the $d=8$ cat state $\ket{\psi^d_{\rm cat}}=(\ket{-7/2}+\ket{7/2})/ \sqrt{2}$, which is maximally sensitive to magnetic $I_z$ noise. 
Although the sequence corrects fluctuations in all diagonal operators, it only corrects noise within a specific frequency band determined by the pulse interval \cite{hansen2023accessing}; high-frequency noise is not corrected for.
Second, adjacent $\Delta m=1$ superpositions are, by their limited dimensionality ($d=2$), already relatively well protected against magnetic and electric noise compared to various other qudit states, as we discuss in the next sections. 

\subsection{Cyclic DD on uniform superpositions and cat states}

Applying the cyclic DD sequence to states whose populations are highly spread out in the $d$-dimensional Hilbert space results in more substantial coherence advantages. 
Here we show cyclic DD on two useful nonclassical states: the uniform superposition state and the Schr\"odinger cat state. The uniform superposition state is spin-squeezed and the cat state displays strong Wigner negativity, both indicators of nonclassicality \cite{Davis2021, Ma2011}. Both spin-squeezed states \cite{Kitagawa1993} and cat states \cite{Chalopin2018} have received broad interest for quantum sensing. 

\begin{figure*}
    \centering
    \includegraphics[width=\linewidth]{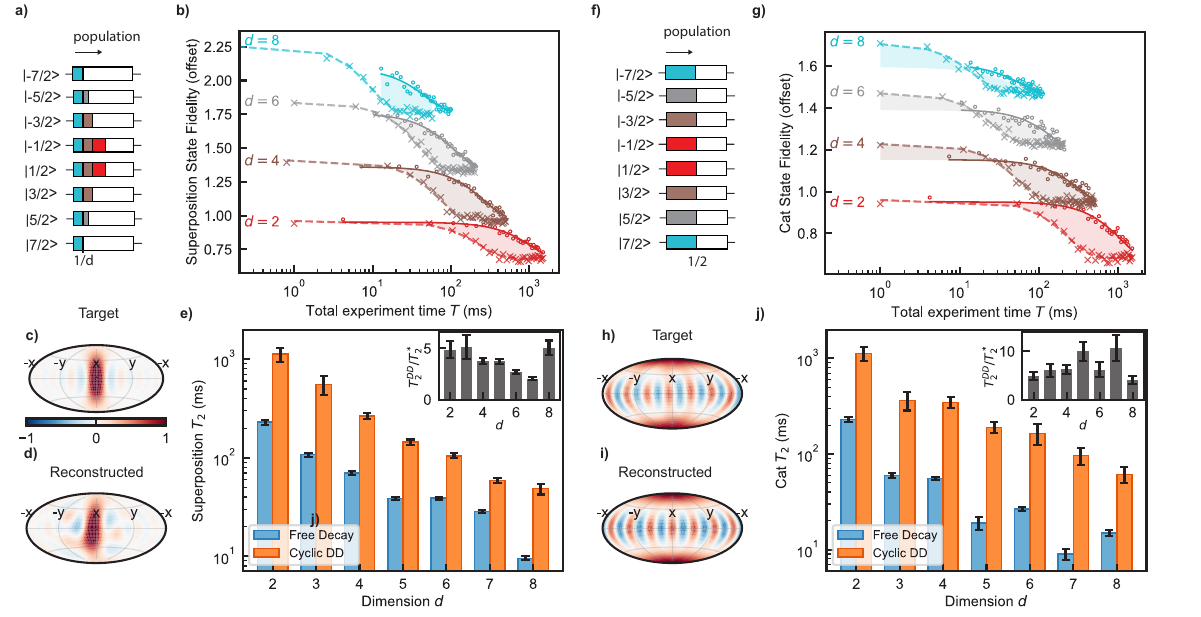}
    \caption{\textbf{Performance of cyclic DD on nonclassical states.}
    \textbf{a)} Overlapping population distributions of uniform superpositions in symmetric, even $d$-dimensional Hilbert spaces. Populations are filled up to $1/d$. Cyan, grey, brown and red correspond to $d=8,6,4,2$ respectively. 
    \textbf{b)} Fidelity decay as a function of the total experiment time $T$ of uniform superposition states for all even dimensions for free decay (x markers) and cyclic DD (o markers). The cyclic DD datapoints start at a later $T$ due to the longer total duration of the $X_d$ operations.
    \textbf{c)} Target and 
    \textbf{d)} reconstructed normalized Wigner function. The color scale is the same for all Wigner functions in this article. ($\mathcal{F}=96.4\%$) of the $d=8$ uniform superposition state at $T=0$. 
    \textbf{e)} Comparison between free decay and cyclic DD dephasing time $T_2$. The inset shows the ratio between $T_2$'s showing up to a factor 5 increase. Odd-dimensional Hilbert spaces are created by including one extra level from the bottom.
    \textbf{f-j)} Same as a-d, but for Schr\"odinger cat states, created with $\mathcal{F}=92.8\%$. 
    }
    
    \label{fig:3}
\end{figure*}

The uniform superposition state
\begin{equation}
\begin{aligned}
    \ket{\psi_{\rm unif}}=\sum_{i=0}^d\frac{1}{\sqrt{d}}\ket{m_{\rm min}+i} \\
    m_{\rm min} =\begin{cases}
-\dfrac{d-1}{2}, & d \text{ even},\\[6pt]
-\dfrac{d-2}{2}, & d \text{ odd},
\end{cases}
\end{aligned}
\end{equation}
with equal population $1/d$ in each of the $d$ states (Fig.~\ref{fig:3}a), is an excellent benchmark for cyclic DD, because the quantum state remains the same after each application of an $X_8$ unitary. The corresponding cyclic DD sequence is implemented by applying $d$ $X_d$ unitaries, which cyclically permute only the states within the chosen subspace. For $d=2$, the sequence simply reduces to the well-known Hahn echo, highlighting cyclic DD as its qudit generalization.

The uniform superposition state is prepared using an arbitrary state initialization protocol \cite{Vaartjes2025} (see Supplementary Information S3). After the DD sequence, the state is read out using density matrix tomography~\cite{Yu2024}, which allows to visualize the spin Wigner function (Fig.~\ref{fig:3}b).
We perform an interleaved comparison measurement of free decay and cyclic DD for various total wait times $T$ (Fig.~\ref{fig:3}d). For all dimensions we observe an increase in $T_2$ up to a factor 5 when using cyclic DD. Because the uniform superposition state is unchanged throughout the whole sequence, the coherence improvement can only result from decoupling of system-bath interactions, rather than an artifact of cycling over particular states.

Next, we perform the same cyclic DD sequence on $d$-dimensional spin cat states. These states are defined as:
\begin{equation}
    \ket{\psi_{\text{cat}}^{d}} = \frac{1}{\sqrt{2}}
    \begin{cases}
    \ket{\tfrac{-d+1}{2}} + \ket{\tfrac{d-1}{2}}, &   d \text{ even}, \\[6pt]
    \ket{\tfrac{-d+2}{2}} + \ket{\tfrac{d}{2}}, &  d \text{  odd},
    \end{cases}
\end{equation}
for \(d \in \{2,3, \cdots,8\}\). These states can be thought of as \(d\)-dimensional embeddings of ``generalized cat states'' inside the eight-dimensional representation of \(SU(2)\) (which represents the total available Hilbert space for the $\Sb$ nuclear spin).

The first iteration of the $X_d$ unitary brings the cat state to a $\Delta m=1$ superposition state. The subsequent $X_d$ unitaries move the superposition one level down, until after $d$ applications, the cycle is completed and we return to the initial cat state. Similar to the uniform superposition states, we see a factor 5 to 10 increase in $T_2$. For the cat states, we see a strong even-odd effect: odd dimensional cat states seem to benefit more from DD. This can qualitatively be understood from the fact that even-$d$ cat states are already insensitive to electric noise coupling through $I_z^2$, whereas odd-$d$ cat states are not. 

\subsection{Cyclic DD on Haar random states}

So far, we verified the benefits of cyclic DD on two specific examples of nonclassical states. This raises the question how the cyclic DD sequence performs on an arbitrary random qudit state. We address this by benchmarking its performance on a set of Haar random states~\cite{anand2026quditdesigns}, i.e. states that uniformly sample the Hilbert space and average to the maximally mixed density matrix.  
In a technical sense, they are the most random quantum states and therefore optimal from the point of view of benchmarking the performance of dynamical decoupling. These states naturally show up in randomized benchmarking, classical shadow tomography, and other quantum information primitives \cite{anand2026quditdesigns}. 

To generate Haar-random states in the qudit Hilbert space, we apply a Haar-random unitary (computed with QuTiP~\cite{Johannson2013}) to any reference state, in this case the \(|7/2\rangle\) state (See Supplementary Information S2 for more details on state initialization). Haar-randomness ensures that the results of benchmarking are independent of the choice of initial reference state.  

\begin{figure}
    \centering
    \includegraphics[width=\linewidth]{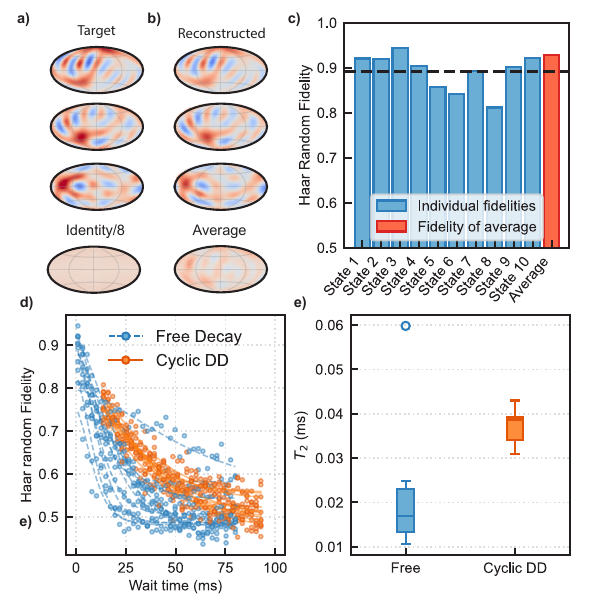}
    \caption{\textbf{Cyclic DD sequence on Haar random states.}
    \textbf{a)} Target and reconstructed Wigner functions of three examples of Haar random states.
    \textbf{b)} The average over all Haar random states should result in the maximally mixed state $\rho=\mathbb{I}/d$. The average reconstructed density matrix is maximally mixed with $\mathcal{F}=92.7\%$. 
    \textbf{c)} Fidelities of 10 created Haar random states (blue, $\mathcal{F}_{\rm avg}=89.2\%$), and fidelity of the averaged state (red).
    \textbf{d)} Fidelity decay of 10 Haar random states for free evolution and cyclic DD.
    \textbf{e)} Box plot of the median $T_2$ in both cases.
    }
    \label{fig:4}
\end{figure}

Figure~\ref{fig:4}a shows the resulting target and reconstructed Wigner functions of three examples. Using density matrix tomography, we measure an average state fidelity of $\mathcal{F}=89.2\%$ with respect to the target states. To verify that the ensemble is indeed (close to) Haar-random, we inspect the average density matrix of the prepared states. For an ideal Haar-random ensemble, this average should result in the maximally mixed state. We confirm this randomness by finding $\mathcal{F}=92.7\%$ with respect to $\rho=\mathbb{I}/8$ (red bar in Fig.~\ref{fig:4}c). 

We perform cyclic DD over the set of states (Fig.~\ref{fig:4}d-e) and measure an increase in the median dephasing time from $T_2^*=16.2$~ms to $T_{2, {\rm cyclic}}=38.6$~ms. 
The coherence time improvement for Haar random states is less pronounced than for highly structured states such as uniform superpositions or cat states since they generally lack symmetries. 

Since the fidelity/overlap is a quadratic function of a quantum state, one can replicate the average fidelity decay curve of a Haar-random initial state by instead averaging over a set of states that form a projective 2-design~\cite{anand2026quditdesigns}. Examples of such states are the stabilizer states of a qubit or, in this case, since the qudit Hilbert space is a prime-power, the stabilizer states for an eight-dimensional qudit. Therefore, the ability to protect Haar-random states under DD acts as a systematic average-case performance over stabilizer states.

\subsection{Comparison of DD techniques on logical cat code states.}

A particularly relevant class of qudit states for quantum information are the logical codewords for the spin-cat code~\cite{Gross2024, Lim2023Fault-tolerantQudit}. The logical states are defined as the spin coherent states oriented along the physical $x$-axis (in a generalized rotating frame): 
\begin{equation}
    \begin{cases}
        \ket{0}_L=\ket{-7/2}_x \\
        \ket{1}_L=\ket{7/2}_x,
    \end{cases}
\end{equation} 
where the subscript $x$ indicates eigenstates of the $I_x$ operator. Therefore, superpositions between these logical states are spin-cat states in the $x$-basis: $\ket{\pm}_{L}=1/\sqrt2\left(\ket{-7/2}_x \pm \ket{7/2}_x\right)$.

In this encoding, the dephasing operators $I_z$ and $I_z^2$ act as coupling operators between $\ket{m_x}$-states: $I_z$ changes $m_x$ by 1, and $I_z^2$ changes $m_x$ by 2.
This means they can induce logical bit-flips, by indirectly coupling the $\ket{+7/2}_x$ and $\ket{-7/2}_x$ states via the intermediate states.
Logical errors occur if the states leak from the positive $m_x$ to the negative $m_x$ subspace, or vice versa~\cite{Lim2023Fault-tolerantQudit, Kusano2026Spin-CatArray}. 

\subsubsection{Quantum error correction and dynamical decoupling}

Over the years, dynamical decoupling has been applied extensively to protect physical qubits and on near-term hardware. However, as it turns out, DD can be optimally combined with quantum error correction (QEC). For example, given a stabilizer code, one can generate a DD sequence from suitable Pauli operators in the normalizer of the stabilizer group such that it preserves the code and its syndrome sectors~\cite{paz2013optimally,Vezvaee2026LogicalDD,kasatkin2026quantum}. As a result, pairing DD with QEC can improve logical performance by reducing threshold requirements~\cite{paz2013optimally,Vezvaee2026LogicalDD,kasatkin2026quantum}. State-of-the-art logical experiments often include simple DD sequences during stabilizer measurements to reduce the effect of dephasing noise~\cite{acharyaSuppressingQuantumErrors2023}.

\subsubsection{Overview of DD strategies}

As opposed to Haar random states or uniform superpositions, the logic cat code may benefit from a decoupling approach tailored to the noise bias and structure of the states. Therefore, besides cyclic DD, we explore three different decoupling strategies: Hahn echo, rotary echo and spin locking, displayed in Fig.\ref{fig:5}b-e. Here we provide an overview of the pulse sequences.

\textbf{Cyclic DD}, as before, involves cyclically permuting all 8 eigenstates of the qudit using the shift operator $X_8$:
    \begin{equation}
        f_{\tau/16}-[X_{8}-f_{\tau/8}]^7-X_8-f_{\tau/16},
    \end{equation}
where $f_\tau$ indicates a period of free evolution.

\textbf{Hahn echo DD}, which decouples magnetic $I_z$ noise (see Supplementary Information S2.B), 
is achieved by two mirrored SU(2)-covariant $R_y(\pi)$ pulses (see Supplementary information S4.C), where each of the $Y$ pulses (see Fig.\ref{fig:5}b) swaps the $x-$basis states:
\begin{equation}
    f_{\tau/4} -R_y(\pi) -f_{\tau/2}-R_{-y}(\pi) -f_{\tau/4}.
\end{equation}
We apply the \textit{SU(2)} Hahn echo with symmetrized pulses to minimize the impact of control errors.

Cyclic and Hahn echo DD sequences rely on 
the ideal pulse limit \cite{Viola1998,viola2003robust}. This can be problematic in experiments like ours, where the pulse time of the \textit{SU(2)}-rotations ($\sim \SI{3.8}{\milli \second}$) or ladder pulses ($\sim \SI{1.5}{\milli \second}$ to execute 7 sequential $\pi$ pulses for one $X_8$ unitary) is non-negligible compared to the dephasing time.

This complication is avoided with the use of continuous DD sequences, which we outline below.

\textbf{Rotary echo DD} has been used in previous studies to cancel quadrupolar shifts in trapped ions \cite{Shaniv2019QuadrupoleDecoupling}. The pulse sequence (Fig.~\ref{fig:5}b, green) cancels quasistatic $I_z^2$ noise terms (see Supplementary Information S6). It can be represented as:
    \begin{equation}
        f_{\tau/6}-R_{y}(\pi/2)-R_{x}(\Omega \tau/3)-R_{-x}(\Omega \tau/3)-R_{-y}(\pi/2)-f_{\tau/6},
    \end{equation}
where $\Omega=2\pi \times \SI{95}{\hertz}$ is the Rabi frequency of the $SU(2)$ rotation  (see Supplementary S7).

\textbf{Spin locking} has a similar pulse sequence, represented by: 
        \begin{equation}
        R_{x}(\Omega \tau/2)-R_{-x}(\Omega \tau/2),
    \end{equation}
where we exploit that the logical states in the cat code are extremal $I_x$ eigenstates. Under a continuous global drive along $I_x$, the quantization axis of the spin becomes $I_x$ in the generalized rotating frame.

By applying the continuous drive in a net-zero pulse, i.e. the rotation reverses after $t=\tau/2$, pulse-induced errors such as over- and under-rotations cancel out.

\subsubsection{Results on logical cat-code states}

\begin{figure*}[thbp]
    \centering
    \includegraphics[width=\linewidth]{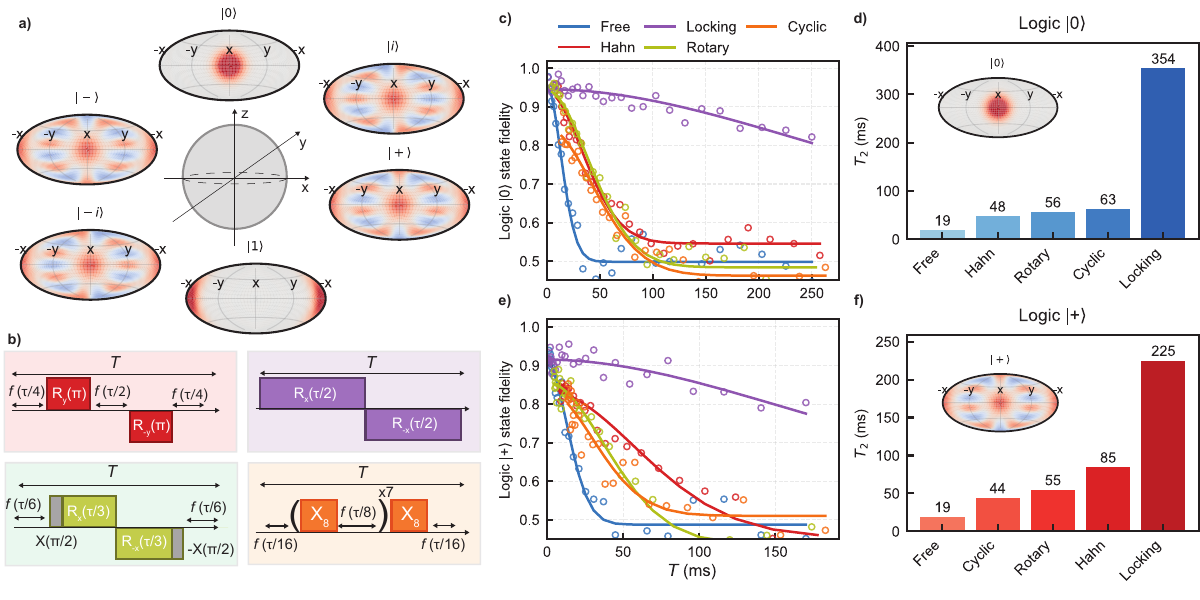}
    \caption{\textbf{Comparison of dynamical decoupling strategies on logical cat code states.}
    \textbf{a)} Wigner functions of the logical code words of the spin-cat code, all displaying the $I_x$-axis as their primary symmetry axis.
    \textbf{b)} Schematic representation of the implemented dynamical decoupling protocols: Hahn echo (red), spin locking (purple), rotary echo (green), cyclic DD (orange).
    \textbf{c)} Measured coherence decay of the logical states $\ket{0}_L$ and 
    \textbf{d)} $\ket{+}_L$ under each DD strategy. 
    \textbf{e-f)} Comparison of all strategies.
    Pulsed sequences (Hahn echo, cyclic) both yield improvements for $\ket{0}_L$ and $\ket{+}_L$. 
    Continuous sequences (rotary, spin locking) provide significant protection, with spin locking achieving the longest coherence times.}
    \label{fig:5}
\end{figure*}

We compare all DD strategies (displayed in Fig.~\ref{fig:5}b-f) applied to the logical cat code states $\ket{0}_L$ and $\ket{+}_L$. The states $\ket{1}_L$ and $\ket{-}_L$ were not measured here, because they only differ by a global $I_z$-rotation of $\pi$, under which the $I_z$ and $I_z^2$ noise operators are invariant. Since the fidelity decay of $\ket{\pm i}_L$ was not measured, the present measurements do not fully characterize the storage of an arbitrary logical qubit. 

The pulsed methods - cyclic DD and Hahn DD - both show coherence improvements on $\ket{0}_L$ and $\ket{+}_L$. Interestingly, the Hahn DD sequence has a stronger effect on $\ket{+}_L$ compared to $\ket{0}_L$ - we measure a factor 5 versus 3 improvement, respectively. 

Among continuous DD techniques, the rotary echo increases the coherence time of both logical states by suppressing quadrupolar $I_z^2$ noise (See Supplementary Information S6), leading to a 2-3$\times$ increase in $T_2$.
We observe the strongest extension of the coherence time using the spin locking sequence with an optimized SU(2) drive amplitude. For the logical superposition $\ket{+}_L$, we obtain $T^{\rm locking}_2 = 225$~ms, compared with approximately $\SI{19}{\milli \second}$ under free evolution and $\SI{85}{\milli \second}$ for a Hahn echo: improvements of about $12\times$ and $2.6\times$. For the logical basis state $\ket{0}_L$, we obtain $T^{\rm locking}_2 = \SI{354}{\milli \second}$.

Spin locking applied to logical spin-cat codewords can be viewed as the high-dimensional generalization of a dressed qubit \cite{Laucht2017ASilicon}. A dressed qubit is only sensitive to noise in $I_z$ or $I_z^2$ centered around the dressing frequency $\Omega$. Since both magnetic ($I_z$) and electric ($I_z^2$) noises have a power spectral density that decreases with frequency (see Supplementary Information S5), their impact on the qubit coherence is reduced by the dressing.

An interesting question for future work is whether one could define a logical qubit in the `dressed-cat' states of the 8-dimensional qudit.

\section{Discussion and Conclusions}

In this work, we demonstrated up to an order-of-magnitude increase of the coherence times of various $d=8$ qudit states such as uniform superposition states and Schr\"odinger cat states. Using the intrinsic noise bias of the system, we employed a subgroup of the universal decoupling group for qudits, that only protects against dominant dephasing errors, thus reducing the number of decoupling unitaries from $d^2$ to $d$. This added efficiency is promising for applications in qudit-based quantum sensing and quantum information processing. Importantly, we further provide an average-case benchmark using Haar-random states spanning the full Hilbert space, demonstrating an increase in the coherence time over ten sampled states.

Additionally, DD offers selective suppression of specific noise terms in the Hamiltonian. Whereas cyclic DD targets all diagonal noise, the rotary echo protocol suppresses only quadratic $I_z^2$ terms. This selective cancellation is relevant for quantum sensing applications, where suppressing unwanted electric quadrupolar shifts ($\propto I_z^2$) can enhance sensitivity to magnetic fields ($\propto I_z$)~\cite{Shaniv2019QuadrupoleDecoupling}.

In the context of quantum computing and quantum error correction, we focused on increasing the coherence times of the logical states of the spin-cat code. By comparing several different decoupling strategies, both stroboscopic and continuous, we found an easy to implement continuous DD method that increases the logical lifetimes from $T_2=\SI{19}{\milli \second}$ to $T_2=\SI{225}{\milli \second}$ for the logical $\ket{+}_L$ state and $T_2=\SI{354}{\milli \second}$ for the logical $\ket{0}_L$ state without performing any active error correction. 

Overall, these results show that DD, especially continuous schemes like spin locking, is a practical and simple way to significantly extend the coherence time of logical spin cat states, which can help reduce the overhead required for active error correction by suppressing noise during idle periods. However, we note that it is not trivial to design a combined DD and QEC protocol, and that several conditions must be met for DD to be beneficial in such a hybrid setting~\cite{paz2013optimally, kasatkin2026quantum}. 

First, the logical DD sequence must not mix syndrome subspaces~\cite{paz2013optimally}. This condition is met since the continuous $I_x$ drive commutes with the syndrome observable $I_x^2$. There is, however, an important consideration for combining \textit{spin locking} and QEC in the spin cat code. 
The spin locking drive introduces logic phase rotations, which either need to be kept track of in a dressed basis~\cite{Laucht2017ASilicon, Hansen2022ImplementationSilicon, Hansen2021PulseComputation}, or inverted using a symmetric net-zero pulse as shown in our implementation in Fig.~\ref{fig:5}. Furthermore, during continuous $I_x$ spin locking, the intrinsic noise bias may be reduced due to control errors introduced by the drive. This motivates further exploration of alternative continuous driving strategies, such as a stabilizing two-photon drive analogous to Kerr-cat qubits~\cite{Grimm2020StabilizationQubit, Puri2020, Ding2025QuantumQubit, Masuda2025StabilizationRefrigerator, McIntyre2026SpinQubits}.

Overall, our results establish dynamical decoupling as a powerful and accessible tool for extending coherence in high-dimensional qudit platforms. These techniques offer a promising path to reduce the overhead of quantum error correction, particularly in biased-noise codes. Developing a unified strategy, i.e., combining qudit DD and QEC, is an important step on the road to building a qudit-based fault-tolerant quantum computer.

\appendix

\section{Materials and Methods}

\subsection{Fabrication and Experimental setup}

The fabrication of the $\Sb$ devices and the experimental setup are described in detail in Ref.~\cite{yu2025schrodinger}. The setup and fabrication of device A and device B are nominally identical, except for the ion implantation parameters. For device A, we used an implantation energy of \SI{10}{\kilo e \volt} and an ion fluence of \SI{4e11}{\centi\meter^{-2}}, whereas for device B, we used \SI{18}{\kilo e \volt}, \SI{5e11}{\centi\meter^{-2}}.

\subsection{Device Operation}
\label{sec:device_operation}

In this work, we perform DD experiments on a single spin-7/2 $\Sb$ donor \cite{Asaad2020, Fernandez2024, Yu2024}, implanted into an isotopically enriched silicon chip \cite{Itoh2014} (Fig.~\ref{fig:1}a). A hyperfine-coupled electron, tunnel-coupled to a Single Electron Transistor (SET), serves as an ancilla for Elzerman readout \cite{Elzerman2004Single-shotDot, Morello2010, Pla2013}, which we perform using an adaptive protocol to minimize measurement backaction \cite{vaartjes2026maximizing}. 
We apply an external magnetic field $\mathbf{B_0}=1.395~\si{\tesla}~\mathbf{\hat{z}}$ in plane with the chip to split the energy levels of both the nucleus and the electron (Fig.~\ref{fig:1}b). The second order energy splitting, equivalent to anharmonicity in superconducting qubits, is provided by the quadrupole interaction which couples electric field gradients (EFG) to the quadratic spin operators (Equation~\ref{eq:system_hamiltonian}). The average quadrupole splitting of $\Delta f_q=\SI{-18.0 \pm 0.9}{\kilo \hertz}$ makes each qudit level individually addressable. 

We apply an oscillating magnetic field $\mathbf{B_1}(t)$ - orthogonal to the external magnetic field - with an on-chip antenna to enable universal control over the qudit via Nuclear Magnetic Resonance (NMR) pulses (see Supplementary Information S4.A). The nuclear qudit control toolbox contains three methods to achieve universal control:\\

\begin{enumerate}
    \item \textbf{Ladder operations} \cite{Cybenko2001}. These operations are activated by a single-tone NMR pulse resonant on a single transition \cite{Asaad2020, Fernandez2024}.
    \item \textbf{SU(2)-covariant rotations}. Rigid rotations in the whole $d=8$ Hilbert space, generated by the $I_x, I_y$ and $I_z$ spin operators.
    We generate these rotations by simultaneously driving all 7 NMR transitions with equal amplitude \cite{Yu2024}.
    \item \textbf{Selective Number Arbitrary Phase (SNAP) gates} \cite{Heeres2015CavityGates}, which apply an arbitrary diagonal unitary to the qudit state. In a Generalized Rotating Frame (GRF) \cite{Leuenberger2003}, SNAP gates can be applied virtually and near-instantaneously by shifting the clock phases of the hardware that keeps track of the GRF \cite{Yu2024}.  
\end{enumerate}

This toolbox allows us to initialize arbitrary states (see Supplementary Information S3), compile any unitary needed for DD (Supplementary Information S4) and perform density matrix tomography~\cite{Yu2024, Vaartjes2025}. 

\subsection*{Lead contact}

Requests for further information and resources should be directed to and will be fulfilled by the lead contact, Andrea Morello (\url{a.morello@unsw.edu.au})

\subsection*{Data and code availability}

All data and code to support the claims in the text can be found in the online repository at \href{https://doi.org/10.5281/zenodo.23074200}{10.5281/zenodo.23074200}

\subsection*{Declaration of interests}

A.M. is an inventor on a patent related to this work, describing the use of high-spin donor nuclei as quantum information processing elements (application no. AU2019227083A1, US16/975,669, WO2019165494A1). A.S.D. is the CEO and a director of Diraq Pty Ltd.. F.E.H. and A.S.D. declare equity interest in Diraq Pty Ltd. All other authors declare no competing interests.

\subsection*{Acknowledgments}

The authors would like to thank Thaddeus Ladd for inspiring discussions about non-Markovian noise.

The research was funded by an Australian Research Council Discovery Project (project no. DP210103769), by the US Army Research Office (Contract no. W911NF-23-1-0113), and by the Australian Department of Industry, Innovation and Science (grant no. AUSMURI000002). A.M. acknowledges support from an Australian Research Council Laureate Fellowship (project no. FL240100181).
We acknowledge the facilities, and the scientific and technical assistance provided by the UNSW node of the Australian National Fabrication Facility (ANFF), and the Heavy Ion Accelerators (HIA) nodes at the University of Melbourne and the Australian National University. ANFF and HIA are supported by the Australian Government through the National Collaborative Research Infrastructure Strategy (NCRIS) program. Ion beam facilities employed by D.N.J. and A.M.J. were co-funded by the Australian Research Council Centre of Excellence for Quantum Computation and Communication Technology (Grant No. CE170100012).
The model fitting and the simulations were partly executed with the assistance of resources from the National Computational Infrastructure (NCI Australia), an NCRIS enabled capability supported by the Australian Government.
A.V., B.W., M.R.v.B., X.Y and L.A.O'N acknowledge support from the Sydney Quantum Academy. D.N.J. acknowledges the support of a Royal Society (UK) Wolfson Visiting Fellowship RSWVF/211016. N.A. is thankful for support from KBR under the Prime Contract No. 80ARC020D0010 with the NASA Ames Research Center. J.M.
is thankful for support from NASA Academic Mission Services, Contract No. NNA16BD14C. The views and conclusions contained in this document are those of the authors and should not be interpreted as representing the official policies, either expressed or implied, of the Army Research Office or the United States Government. The United States Government retains, and by accepting the article for publication, the publisher acknowledges that the United States Government retains, a nonexclusive, paid-up,
irrevocable, worldwide license to publish or reproduce the published form of this work, or allow
others to do so, for United States Government purposes.

\subsection*{Author contributions}   

A.V. conducted the experiments and data analysis with input from R.Y.S., B.W., M.R.v.B., X.Y. and M.N. 
N.A., J.M. developed the theory and cyclic DD pulse sequences, with input from B.P..
R.Y.S. and M.N. developed the measurement software.
A.Y. and R.S. built the non-Markovian noise model and performed model simulations. D.H. and F.E.H. fabricated the device with A.S.D.’s supervision. 
A.V., B.W., M.R.v.B., and X.Y. performed device characterization and tune-up. 
L.A.O'N performed device simulations. 
D.H. and A.M.J performed the ion implantation with D.N.J's supervision. K.M.I. supplied the isotopically enriched silicon. A.V., N.A., A.Y. and A.M. wrote the manuscript with input from the other authors. A.M. supervised the project.

\bibliography{bib}

@article{PhysRevA.92.040303,
  title = {Universal control of an oscillator with dispersive coupling to a qubit},
  author = {Krastanov, Stefan and Albert, Victor V. and Shen, Chao and Zou, Chang-Ling and Heeres, Reinier W. and Vlastakis, Brian and Schoelkopf, Robert J. and Jiang, Liang},
  journal = {Phys. Rev. A},
  volume = {92},
  issue = {4},
  pages = {040303(R)},
  numpages = {5},
  year = {2015},
  month = {Oct},
  publisher = {American Physical Society},
  doi = {10.1103/PhysRevA.92.040303},
  url = {https://link.aps.org/doi/10.1103/PhysRevA.92.040303}
}

@article{sun2024permutational,
  title = {Full-Permutation Dynamical Decoupling in Triple-Quantum-Dot Spin Qubits},
  author = {Sun, Bo and Brecht, Teresa and Fong, Bryan H. and Akmal, Moonmoon and Blumoff, Jacob Z. and Cain, Tyler A. and Carter, Faustin W. and Finestone, Dylan H. and Fireman, Micha N. and Ha, Wonill and Hatke, Anthony T. and Hickey, Ryan M. and Jackson, Clayton A. C. and Jenkins, Ian and Jones, Aaron M. and Pan, Andrew and Ward, Daniel R. and Weinstein, Aaron J. and Whiteley, Samuel J. and Williams, Parker and Borselli, Matthew G. and Rakher, Matthew T. and Ladd, Thaddeus D.},
  journal = {PRX Quantum},
  volume = {5},
  issue = {2},
  pages = {020356},
  numpages = {20},
  year = {2024},
  month = {Jun},
  publisher = {American Physical Society},
  doi = {10.1103/PRXQuantum.5.020356},
  url = {https://link.aps.org/doi/10.1103/PRXQuantum.5.020356}
}

@article{PhysRevA.71.052318,
  title = {Criteria for exact qudit universality},
  author = {Brennen, Gavin K. and O'Leary, Dianne P. and Bullock, Stephen S.},
  journal = {Phys. Rev. A},
  volume = {71},
  issue = {5},
  pages = {052318},
  numpages = {7},
  year = {2005},
  month = {May},
  publisher = {American Physical Society},
  doi = {10.1103/PhysRevA.71.052318},
  url = {https://link.aps.org/doi/10.1103/PhysRevA.71.052318}
}

@article{read2025groupfactorization,
  title = {Dynamical decoupling of interacting spins through group factorization},
  volume = {112},
  ISSN = {2469-9934},
  url = {http://dx.doi.org/10.1103/w5f7-4z5l},
  DOI = {10.1103/w5f7-4z5l},
  number = {4},
  journal = {Physical Review A},
  publisher = {American Physical Society (APS)},
  author = {Read,  Colin and Serrano-Ensástiga,  Eduardo and Martin,  John},
  year = {2025},
  pages = {042601},
  month = Oct 
}

@article{nielsen_universal_2002,
   title={Universal simulation of Hamiltonian dynamics for quantum systems with finite-dimensional state spaces},
   volume={66},
   ISSN={1094-1622},
   url={http://dx.doi.org/10.1103/PhysRevA.66.022317},
   DOI={10.1103/physreva.66.022317},
   number={2},
   journal={Physical Review A},
   publisher={American Physical Society (APS)},
   author={Nielsen, Michael A. and Bremner, Michael J. and Dodd, Jennifer L. and Childs, Andrew M. and Dawson, Christopher M.},
   year={2002},
   pages={022317},
   month=Aug }

@article{Vezvaee2026LogicalDD,
    author = {Vezvaee, Arian and Tripathi, Vinay and Morford-Oberst, Mario and Butt, Friederike and Kasatkin, Victor and Lidar, Daniel A.},
    title = {Demonstration of high-fidelity entangled logical qubits using transmons},
    journal = {Nature Communications},
    volume = {17},
    number = {1},
    pages = {1516},
    year = {2026},
    month = {2},
    doi = {10.1038/s41467-026-70011-3}
}

@article{anand2026quditdesigns,
      title={{Qudit Designs and Where to Find Them}}, 
      author={Namit Anand and Jeffrey Marshall and Jason Saied and Eleanor Rieffel and Andrea Morello},
      year={2026},
      journal = {arXiv preprint},
      eprint={2603.02659}
}

@article{Tripathi2022,
    author = {Vinay Tripathi and Huo Chen and Mostafa Khezri and Ka Wa Yip and E. M. Levenson-Falk and Daniel A. Lidar},
    title = {Suppression of Crosstalk in Superconducting Qubits Using Dynamical Decoupling},
    journal = {Physical Review Applied},
    volume = {18},
    number = {2},
    pages = {024068},
    year = {2022},
    doi = {10.1103/PhysRevApplied.18.024068}
}

@article{Han2025,
    author = {Jia-Xiu Han and Jiang Zhang and Guang-Ming Xue and Haifeng Yu and Guilu Long},
    title = {Protecting logical qubits with dynamical decoupling},
    journal = {Physical Review Applied},
    volume = {24},
    number = {2},
    pages = {024068},
    year = {2025},
    doi = {10.1103/lm8x-r48q}
}

@article{ezzell2023dynamical,
  title={Dynamical decoupling for superconducting qubits: A performance survey},
  author={Ezzell, Nic and Pokharel, Bibek and Tewala, Lina and Quiroz, Gregory and Lidar, Daniel A},
  journal={Physical Review Applied},
  volume={20},
  number={6},
  pages={064027},
  year={2023},
  publisher={APS},
  doi = {10.1103/PhysRevApplied.20.064027}
}

@article{Uhrig2011,
   author = {Götz S. Uhrig},
   doi = {10.1088/1367-2630/13/5/059504},
   issn = {13672630},
   journal = {New Journal of Physics},
   month = {5},
   title = {Exact results on dynamical decoupling by $\pi$ pulses in quantum information processes},
   volume = {13},
   year = {2011},
   pages={059504}
}

@article{lidar_lecture_2020,
      title={Lecture Notes on the Theory of Open Quantum Systems}, 
      author={Daniel A. Lidar},
      year={2020},
        journal = {arXiv preprint},
      eprint={1902.00967}

}

@Inbook{Chizhik2014,
author="Chizhik, Vladimir I.
and Chernyshev, Yuri S.
and Donets, Alexey V.
and Frolov, Vyacheslav V.
and Komolkin, Andrei V.
and Shelyapina, Marina G.",
title="Nuclear Quadrupole Resonance",
bookTitle="Magnetic Resonance and Its Applications",
year="2014",
publisher="Springer International Publishing",
address="Cham",
pages="415--479",
isbn="978-3-319-05299-1",
doi="10.1007/978-3-319-05299-1_8",
url="https://doi.org/10.1007/978-3-319-05299-1_8"
}

@article{Gullion1990,
title = {New, compensated Carr-Purcell sequences},
journal = {Journal of Magnetic Resonance (1969)},
volume = {89},
number = {3},
pages = {479-484},
year = {1990},
issn = {0022-2364},
doi = {https://doi.org/10.1016/0022-2364(90)90331-3},
url = {https://www.sciencedirect.com/science/article/pii/0022236490903313},
author = {Terry Gullion and David B Baker and Mark S Conradi}
}

@article{Viola1998,
  title = {Dynamical suppression of decoherence in two-state quantum systems},
  author = {Viola, Lorenza and Lloyd, Seth},
  journal = {Phys. Rev. A},
  volume = {58},
  issue = {4},
  pages = {2733--2744},
  numpages = {0},
  year = {1998},
  month = {Oct},
  publisher = {American Physical Society},
  doi = {10.1103/PhysRevA.58.2733},
  url = {https://link.aps.org/doi/10.1103/PhysRevA.58.2733}
}

@article{Ajoy2011,
    author = {Ashok Ajoy and Gonzalo A. Álvarez and Dieter Suter},
    title = {Optimal pulse spacing for dynamical decoupling in the presence of a purely dephasing spin bath},
    journal = {Physical Review A},
    volume = {83},
    number = {3},
    year = {2011},
    pages = {032303},
    doi = {10.1103/PhysRevA.83.032303}
}

@article{Tripathi2025characterization,
   author = {Vinay Tripathi and Daria Kowsari and Kumar Saurav and Haimeng Zhang and Eli M. Levenson-Falk and Daniel A. Lidar},
   doi = {10.1021/acs.chemrev.4c00870},
   issn = {15206890},
   issue = {12},
   journal = {Chemical Reviews},
   month = {6},
   pages = {5745-5775},
   pmid = {40324106},
   publisher = {American Chemical Society},
   title = {Benchmarking Quantum Gates and Circuits},
   volume = {125},
   year = {2025}
}

@article{
DeLange2010,
author = {G. de Lange  and Z. H. Wang  and D. Ristè  and V. V. Dobrovitski  and R. Hanson },
title = {Universal Dynamical Decoupling of a Single Solid-State Spin from a Spin Bath},
journal = {Science},
volume = {330},
number = {6000},
pages = {60-63},
year = {2010},
doi = {10.1126/science.1192739}}

@article{Itoh2014,
   author = {Kohei M. Itoh and Hideyuki Watanabe},
   doi = {10.1557/mrc.2014.32},
   issn = {21596867},
   issue = {4},
   journal = {MRS Communications},
   month = {9},
   pages = {143-157},
   publisher = {Materials Research Society},
   title = {Isotope engineering of silicon and diamond for quantum computing and sensing applications},
   volume = {4},
   year = {2014}
}

@article{Savytskyy2023,
author = {Rostyslav Savytskyy  and Tim Botzem  and Irene Fernandez de Fuentes  and Benjamin Joecker  and Jarryd J. Pla  and Fay E. Hudson  and Kohei M. Itoh  and Alexander M. Jakob  and Brett C. Johnson  and David N. Jamieson  and Andrew S. Dzurak  and Andrea Morello },
title = {An electrically driven single-atom “flip-flop” qubit},
journal = {Science Advances},
volume = {9},
number = {6},
pages = {eadd9408},
year = {2023},
doi = {10.1126/sciadv.add9408},
URL = {https://www.science.org/doi/abs/10.1126/sciadv.add9408}}

@article{Asaad2020,
  title = {Coherent Electrical Control of a Single High-Spin Nucleus in Silicon},
  author = {Asaad, Serwan and Mourik, Vincent and Joecker, Benjamin and Johnson, Mark A. I. and Baczewski, Andrew D. and Firgau, Hannes R. and Madzik, Mateusz T. and Schmitt, Vivien and Pla, Jarryd J. and Hudson, Fay E. and Itoh, Kohei M. and McCallum, Jeffrey C. and Dzurak, Andrew S. and Laucht, Arne and Morello, Andrea},
  year = {2020},
  month = mar,
  journal = {Nature},
  volume = {579},
  number = {7798},
  pages = {205--209},
  publisher = {Nature Publishing Group},
  issn = {1476-4687},
  doi = {10.1038/s41586-020-2057-7},
  urldate = {2024-10-09},
  copyright = {2020 This is a U.S. government work and not under copyright protection in the U.S.; foreign copyright protection may apply},
  langid = {english}
}

@article{Chalopin2018,
  title = {Quantum-Enhanced Sensing Using Non-Classical Spin States of a Highly Magnetic Atom},
  author = {Chalopin, Thomas and Bouazza, Chayma and Evrard, Alexandre and Makhalov, Vasiliy and Dreon, Davide and Dalibard, Jean and Sidorenkov, Leonid A. and Nascimbene, Sylvain},
  year = {2018},
  month = nov,
  journal = {Nature Communications},
  volume = {9},
  number = {1},
  pages = {4955},
  publisher = {Nature Publishing Group},
  issn = {2041-1723},
  doi = {10.1038/s41467-018-07433-1},
  urldate = {2024-10-09},
  copyright = {2018 The Author(s)},
  langid = {english}
}

@article{Champion2024,
  title = {Efficient Control of a Transmon Qudit Using Effective Spin-$7/2$ Rotations},
  author = {Champion, Elizabeth and Wang, Zihao and Parker, Rayleigh W. and Blok, Machiel S.},
  journal = {Physical Review X},
  volume = {15},
  issue = {2},
  pages = {021096},
  numpages = {27},
  year = {2025},
  month = {Jun},
  publisher = {American Physical Society},
  doi = {10.1103/vbh4-lysv},
  url = {https://link.aps.org/doi/10.1103/vbh4-lysv}
}

@article{Cybenko2001,
  title = {Reducing Quantum Computations to Elementary Unitary Operations},
  author = {Cybenko, George},
  year = {2001},
  month = apr,
  journal = {Computing in Science \& Engineering},
  volume = {3},
  pages = {27--32},
  doi = {10.1109/5992.908999}
}

@article{Davis2021,
  title = {Wigner Negativity in Spin- j Systems},
  author = {Davis, Jack and Kumari, Meenu and Mann, Robert B. and Ghose, Shohini},
  year = {2021},
  month = aug,
  journal = {Physical Review Research},
  volume = {3},
  number = {3},
  pages = {033134},
  issn = {2643-1564},
  doi = {10.1103/PhysRevResearch.3.033134},
  urldate = {2024-08-12},
  langid = {english}
}

@article{Fernandez2024,
  title = {Navigating the 16-Dimensional {{Hilbert}} Space of a High-Spin Donor Qudit with Electric and Magnetic Fields},
  author = {{Fern{\'a}ndez de Fuentes}, Irene and Botzem, Tim and Johnson, Mark A. I. and Vaartjes, Arjen and Asaad, Serwan and Mourik, Vincent and Hudson, Fay E. and Itoh, Kohei M. and Johnson, Brett C. and Jakob, Alexander M. and McCallum, Jeffrey C. and Jamieson, David N. and Dzurak, Andrew S. and Morello, Andrea},
  year = {2024},
  month = feb,
  journal = {Nature Communications},
  volume = {15},
  number = {1},
  pages = {1380},
  publisher = {Nature Publishing Group},
  issn = {2041-1723},
  doi = {10.1038/s41467-024-45368-y},
  urldate = {2024-10-09},
  copyright = {2024 The Author(s)},
  langid = {english}
}

@article{Vaartjes2025,
    author = {Arjen Vaartjes and Martin Nurizzo and Lin Htoo Zaw and Benjamin Wilhelm and Xi Yu and Danielle Holmes and Daniel Schwienbacher and Anders Kringhøj and Mark R. van Blankenstein and Alexander M. Jakob and Fay E. Hudson and Kohei M. Itoh and Riley J. Murray and Robin Blume-Kohout and Namit Anand and Andrew S. Dzurak and David N. Jamieson and Valerio Scarani and Andrea Morello},
    title = {Certifying the quantumness of a nuclear spin qudit through its uniform precession},
    journal = {Newton},
    volume = {1},
    number = {1},
    pages = {100017},
    year = {2025},
    doi = {10.1016/j.newton.2025.100017}
}

@article{Gross2024,
  title = {Hardware-Efficient Error-Correcting Codes for Large Nuclear Spins},
  author = {Gross, Jonathan A. and Godfrin, Cl{\'e}ment and Blais, Alexandre and {Dupont-Ferrier}, Eva},
  year = {2024},
  month = jul,
  journal = {Physical Review Applied},
  volume = {22},
  number = {1},
  pages = {014006},
  publisher = {American Physical Society},
  doi = {10.1103/PhysRevApplied.22.014006},
  urldate = {2024-10-09}
}

@article{Johannson2013,
  title = {{{QuTiP}} 2: {{A Python}} Framework for the Dynamics of Open Quantum Systems},
  shorttitle = {{{QuTiP}} 2},
  author = {Johansson, J.R. and Nation, P.D. and Nori, Franco},
  year = {2013},
  month = apr,
  journal = {Computer Physics Communications},
  volume = {184},
  number = {4},
  pages = {1234--1240},
  issn = {00104655},
  doi = {10.1016/j.cpc.2012.11.019},
  urldate = {2024-10-09},
  copyright = {https://www.elsevier.com/tdm/userlicense/1.0/},
  langid = {english}
}

@article{Kitagawa1993,
  title = {Squeezed Spin States},
  author = {Kitagawa, Masahiro and Ueda, Masahito},
  year = {1993},
  month = jun,
  journal = {Physical Review A},
  volume = {47},
  number = {6},
  pages = {5138--5143},
  publisher = {American Physical Society},
  doi = {10.1103/PhysRevA.47.5138},
  urldate = {2024-10-09}
}

@article{Leuenberger2003,
  title = {Grover Algorithm for Large Nuclear Spins in Semiconductors},
  author = {Leuenberger, Michael N. and Loss, Daniel},
  year = {2003},
  month = oct,
  journal = {Physical Review B},
  volume = {68},
  number = {16},
  pages = {165317},
  publisher = {American Physical Society},
  doi = {10.1103/PhysRevB.68.165317},
  urldate = {2024-10-09}
}

@article{Morello2010,
  title = {Single-Shot Readout of an Electron Spin in Silicon},
  author = {Morello, Andrea and Pla, Jarryd J. and Zwanenburg, Floris A. and Chan, Kok W. and Tan, Kuan Y. and Huebl, Hans and M{\"o}tt{\"o}nen, Mikko and Nugroho, Christopher D. and Yang, Changyi and {van Donkelaar}, Jessica A. and Alves, Andrew D. C. and Jamieson, David N. and Escott, Christopher C. and Hollenberg, Lloyd C. L. and Clark, Robert G. and Dzurak, Andrew S.},
  year = {2010},
  month = oct,
  journal = {Nature},
  volume = {467},
  number = {7316},
  pages = {687--691},
  publisher = {Nature Publishing Group},
  issn = {1476-4687},
  doi = {10.1038/nature09392},
  urldate = {2024-10-09},
  copyright = {2010 Springer Nature Limited},
  langid = {english}
}

@article{Neeley2009,
  title = {Emulation of a {{Quantum Spin}} with a {{Superconducting Phase Qudit}}},
  author = {Neeley, Matthew and Ansmann, Markus and Bialczak, Radoslaw C. and Hofheinz, Max and Lucero, Erik and O'Connell, Aaron D. and Sank, Daniel and Wang, Haohua and Wenner, James and Cleland, Andrew N. and Geller, Michael R. and Martinis, John M.},
  year = {2009},
  month = aug,
  journal = {Science},
  volume = {325},
  number = {5941},
  pages = {722--725},
  publisher = {American Association for the Advancement of Science},
  doi = {10.1126/science.1173440},
  urldate = {2024-10-09}
}

@article{Pla2013,
  title = {High-Fidelity Readout and Control of a Nuclear Spin Qubit in Silicon},
  author = {Pla, Jarryd J. and Tan, Kuan Y. and Dehollain, Juan P. and Lim, Wee H. and Morton, John J. L. and Zwanenburg, Floris A. and Jamieson, David N. and Dzurak, Andrew S. and Morello, Andrea},
  year = {2013},
  month = apr,
  journal = {Nature},
  volume = {496},
  number = {7445},
  pages = {334--338},
  publisher = {Nature Publishing Group},
  issn = {1476-4687},
  doi = {10.1038/nature12011},
  urldate = {2024-10-09},
  copyright = {2013 Springer Nature Limited},
  langid = {english}
}

@article{Puri2020,
  title = {Bias-Preserving Gates with Stabilized Cat Qubits},
  author = {Puri, Shruti and {St-Jean}, Lucas and Gross, Jonathan A. and Grimm, Alexander and Frattini, Nicholas E. and Iyer, Pavithran S. and Krishna, Anirudh and Touzard, Steven and Jiang, Liang and Blais, Alexandre and Flammia, Steven T. and Girvin, S. M.},
  year = {2020},
  month = aug,
  journal = {Science Advances},
  volume = {6},
  number = {34},
  pages = {eaay5901},
  publisher = {American Association for the Advancement of Science},
  doi = {10.1126/sciadv.aay5901},
  urldate = {2024-10-01}
}

@article{Roy2024,
  title = {Synthetic High Angular Momentum Spin Dynamics in a Microwave Oscillator},
  author = {Roy, Saswata and Senanian, Alen and Wang, Christopher S. and Wetherbee, Owen C. and Zhang, Luojia and Cole, B. and Larson, C. P. and Yelton, E. and Arora, Kartikeya and McMahon, Peter L. and Plourde, B. L. T. and Royer, Baptiste and Fatemi, Valla},
  journal = {Phys. Rev. X},
  volume = {15},
  issue = {2},
  pages = {021009},
  numpages = {24},
  year = {2025},
  month = {Apr},
  publisher = {American Physical Society},
  doi = {10.1103/PhysRevX.15.021009},
  url = {https://link.aps.org/doi/10.1103/PhysRevX.15.021009}
}

@article{Yu2024,
  title = {Schr\"{o}dinger cat states of a nuclear spin qudit in silicon},
  volume = {21},
  ISSN = {1745-2481},
  url = {http://dx.doi.org/10.1038/s41567-024-02745-0},
  DOI = {10.1038/s41567-024-02745-0},
  number = {3},
  journal = {Nature Physics},
  publisher = {Springer Science and Business Media LLC},
  author = {Yu,  Xi and Wilhelm,  Benjamin and Holmes,  Danielle and Vaartjes,  Arjen and Schwienbacher,  Daniel and Nurizzo,  Martin and Kringhøj,  Anders and Blankenstein,  Mark R. van and Jakob,  Alexander M. and Gupta,  Pragati and Hudson,  Fay E. and Itoh,  Kohei M. and Murray,  Riley J. and Blume-Kohout,  Robin and Ladd,  Thaddeus D. and Anand,  Namit and Dzurak,  Andrew S. and Sanders,  Barry C. and Jamieson,  David N. and Morello,  Andrea},
  year = {2025},
  month = jan,
  pages = {362–367},
}

@article{godfrinOperatingQuantumStates2017,
  title = {Operating {{Quantum States}} in {{Single Magnetic Molecules}}: {{Implementation}} of {{Grover}}'s {{Quantum Algorithm}}},
  shorttitle = {Operating {{Quantum States}} in {{Single Magnetic Molecules}}},
  author = {Godfrin, C. and Ferhat, A. and Ballou, R. and Klyatskaya, S. and Ruben, M. and Wernsdorfer, W. and Balestro, F.},
  year = {2017},
  month = nov,
  journal = {Physical Review Letters},
  volume = {119},
  number = {18},
  pages = {187702},
  publisher = {American Physical Society},
  doi = {10.1103/PhysRevLett.119.187702},
  urldate = {2024-11-14}
}

@article{Elzerman2004Single-shotDot,
  title = {Single-shot read-out of an individual electron spin in a quantum dot},
  volume = {430},
  ISSN = {1476-4687},
  url = {http://dx.doi.org/10.1038/nature02693},
  DOI = {10.1038/nature02693},
  number = {6998},
  journal = {Nature},
  publisher = {Springer Science and Business Media LLC},
  author = {Elzerman,  J. M. and Hanson,  R. and Willems van Beveren,  L. H. and Witkamp,  B. and Vandersypen,  L. M. K. and Kouwenhoven,  L. P.},
  year = {2004},
  month = jul,
  pages = {431–435}
}

@article{Shaniv2019QuadrupoleDecoupling,
    author = {Shaniv, Ravid and Akerman, Nitzan and Manovitz, Tom and Shapira, Yotam and Ozeri, Roee},
    title = {Quadrupole Shift Cancellation Using Dynamic Decoupling},
    journal = {Physical Review Letters},
    volume = {122},
    number = {22},
    pages = {223204},
    year = {2019},
    doi = {10.1103/PhysRevLett.122.223204}
}

@article{Laucht2017ASilicon,
    title = {{A dressed spin qubit in silicon}},
    year = {2017},
    journal = {Nature Nanotechnology},
    author = {Laucht, Arne and Kalra, Rachpon and Simmons, Stephanie and Dehollain, Juan P. and Muhonen, Juha T. and Mohiyaddin, Fahd A. and Freer, Solomon and Hudson, Fay E. and Itoh, Kohei M. and Jamieson, David N. and McCallum, Jeffrey C. and Dzurak, Andrew S. and Morello, A.},
    number = {1},
    month = {1},
    pages = {61--66},
    volume = {12},
    publisher = {Nature Publishing Group},
    doi = {10.1038/nnano.2016.178},
    issn = {17483395},
    pmid = {27749833},
    arxivId = {1603.04800}
}

@article{Hansen2021PulseComputation,
    title = {{Pulse engineering of a global field for robust and universal quantum computation}},
    year = {2021},
    journal = {Physical Review A},
    author = {Hansen, Ingvild and Seedhouse, Amanda E. and Saraiva, Andre and Laucht, Arne and Dzurak, Andrew S. and Yang, Chih Hwan},
    number = {6},
    month = {12},
    volume = {104},
    publisher = {American Physical Society},
    pages = {062415},
    doi = {10.1103/PhysRevA.104.062415}
}

@article{Hansen2022ImplementationSilicon,
    title = {{Implementation of an advanced dressing protocol for global qubit control in silicon}},
    year = {2022},
    journal = {Applied Physics Reviews},
    author = {Hansen, I. and Seedhouse, A. E. and Chan, K. W. and Hudson, F. E. and Itoh, K. M. and Laucht, A. and Saraiva, A. and Yang, C. H. and Dzurak, A. S.},
    number = {3},
    month = {9},
    volume = {9},
    publisher = {American Institute of Physics Inc.},
    doi = {10.1063/5.0096467},
    pages = {031409}
}

@article{Tripathi2025QuditProcessor,
    title = {Qudit Dynamical Decoupling on a Superconducting Quantum Processor},
    year = {2025},
    journal = {Physical Review Letters},
    author = {Tripathi, Vinay and Goss, Noah and Vezvaee, Arian and Nguyen, Long B. and Siddiqi, Irfan and Lidar, Daniel A.},
    number = {5},
    month = {2},
    volume = {134},
    pages = {050601},
    publisher = {American Physical Society},
    doi = {10.1103/PhysRevLett.134.050601}
}

@article{Heeres2015CavityGates,
    title = {Cavity State Manipulation Using Photon-Number Selective Phase Gates},
    year = {2015},
    journal = {Physical Review Letters},
    author = {Heeres, Reinier W. and Vlastakis, Brian and Holland, Eric and Krastanov, Stefan and Albert, Victor V. and Frunzio, Luigi and Jiang, Liang and Schoelkopf, Robert J.},
    number = {13},
    month = {9},
    volume = {115},
    pages = {137002},
    publisher = {American Physical Society},
    doi = {10.1103/PhysRevLett.115.137002},
    issn = {10797114},
    arxivId = {1503.01496}
}

@article{Ma2011,
    title = {{Quantum spin squeezing}},
    year = {2011},
    journal = {Physics Reports},
    author = {Ma, Jian and Wang, Xiaoguang and Sun, C. P. and Nori, Franco},
    number = {2-3},
    pages = {89--165},
    volume = {509},
    doi = {10.1016/j.physrep.2011.08.003},
    issn = {03701573},
    arxivId = {1011.2978}
}

@article{paz2013optimally,
  title={Optimally combining dynamical decoupling and quantum error correction},
  author={Paz-Silva, Gerardo A and Lidar, Daniel A},
  journal={Scientific reports},
  volume={3},
  number={1},
  pages={1530},
  year={2013},
  publisher={Nature Publishing Group UK London},
  doi = {10.1038/srep01530}
}

@article{acharyaSuppressingQuantumErrors2023,
  title = {Suppressing Quantum Errors by Scaling a Surface Code Logical Qubit},
  author = {Acharya, Rajeev and Aleiner, Igor and Allen, Richard and Andersen, Trond I. and Ansmann, Markus and Arute, Frank and Arya, Kunal and Asfaw, Abraham and Atalaya, Juan and Babbush, Ryan and Bacon, Dave and Bardin, Joseph C. and Basso, Joao and Bengtsson, Andreas and Boixo, Sergio and Bortoli, Gina and Bourassa, Alexandre and Bovaird, Jenna and Brill, Leon and Broughton, Michael and Buckley, Bob B. and Buell, David A. and Burger, Tim and Burkett, Brian and Bushnell, Nicholas and Chen, Yu and Chen, Zijun and Chiaro, Ben and Cogan, Josh and Collins, Roberto and Conner, Paul and Courtney, William and Crook, Alexander L. and Curtin, Ben and Debroy, Dripto M. and Del Toro Barba, Alexander and Demura, Sean and Dunsworth, Andrew and Eppens, Daniel and Erickson, Catherine and Faoro, Lara and Farhi, Edward and Fatemi, Reza and Flores Burgos, Leslie and Forati, Ebrahim and Fowler, Austin G. and Foxen, Brooks and Giang, William and Gidney, Craig and Gilboa, Dar and Giustina, Marissa and Grajales Dau, Alejandro and Gross, Jonathan A. and Habegger, Steve and Hamilton, Michael C. and Harrigan, Matthew P. and Harrington, Sean D. and Higgott, Oscar and Hilton, Jeremy and Hoffmann, Markus and Hong, Sabrina and Huang, Trent and Huff, Ashley and Huggins, William J. and Ioffe, Lev B. and Isakov, Sergei V. and Iveland, Justin and Jeffrey, Evan and Jiang, Zhang and Jones, Cody and Juhas, Pavol and Kafri, Dvir and Kechedzhi, Kostyantyn and Kelly, Julian and Khattar, Tanuj and Khezri, Mostafa and Kieferov{\'a}, M{\'a}ria and Kim, Seon and Kitaev, Alexei and Klimov, Paul V. and Klots, Andrey R. and Korotkov, Alexander N. and Kostritsa, Fedor and Kreikebaum, John Mark and Landhuis, David and Laptev, Pavel and Lau, Kim-Ming and Laws, Lily and Lee, Joonho and Lee, Kenny and Lester, Brian J. and Lill, Alexander and Liu, Wayne and Locharla, Aditya and Lucero, Erik and Malone, Fionn D. and Marshall, Jeffrey and Martin, Orion and McClean, Jarrod R. and McCourt, Trevor and McEwen, Matt and Megrant, Anthony and Meurer Costa, Bernardo and Mi, Xiao and Miao, Kevin C. and Mohseni, Masoud and Montazeri, Shirin and Morvan, Alexis and Mount, Emily and Mruczkiewicz, Wojciech and Naaman, Ofer and Neeley, Matthew and Neill, Charles and Nersisyan, Ani and Neven, Hartmut and Newman, Michael and Ng, Jiun How and Nguyen, Anthony and Nguyen, Murray and Niu, Murphy Yuezhen and O'Brien, Thomas E. and Opremcak, Alex and Platt, John and Petukhov, Andre and Potter, Rebecca and Pryadko, Leonid P. and Quintana, Chris and Roushan, Pedram and Rubin, Nicholas C. and Saei, Negar and Sank, Daniel and Sankaragomathi, Kannan and Satzinger, Kevin J. and Schurkus, Henry F. and Schuster, Christopher and Shearn, Michael J. and Shorter, Aaron and Shvarts, Vladimir and Skruzny, Jindra and Smelyanskiy, Vadim and Smith, W. Clarke and Sterling, George and Strain, Doug and Szalay, Marco and Torres, Alfredo and Vidal, Guifre and Villalonga, Benjamin and Vollgraff Heidweiller, Catherine and White, Theodore and Xing, Cheng and Yao, Z. Jamie and Yeh, Ping and Yoo, Juhwan and Young, Grayson and Zalcman, Adam and Zhang, Yaxing and Zhu, Ningfeng and {Google Quantum AI}},
  year = 2023,
  journal = {Nature},
  volume = {614},
  number = {7949},
  pages = {676--681},
  publisher = {Nature Publishing Group},
  issn = {1476-4687},
  doi = {10.1038/s41586-022-05434-1},
  copyright = {2023 The Author(s)}
}

@article{tong2025empirical,
  title={Empirical learning of dynamical decoupling on quantum processors},
  author={Tong, Christopher and Zhang, Helena and Pokharel, Bibek},
  journal={PRX Quantum},
  volume={6},
  number={3},
  pages={030319},
  year={2025},
  publisher={APS},
  doi={10.1103/h7pq-s159}
}

@article{coote2025resource,
  title={Resource-efficient context-aware dynamical decoupling embedding for arbitrary large-scale quantum algorithms},
  author={Coote, Paul and Dimov, Roman and Maity, Smarak and Hartnett, Gavin S and Biercuk, Michael J and Baum, Yuval},
  journal={PRX Quantum},
  volume={6},
  number={1},
  pages={010332},
  year={2025},
  publisher={APS},
  doi={10.1103/PRXQuantum.6.010332}
}

@article{Muhonen2014,
    title = {{Storing quantum information for 30 seconds in a nanoelectronic device}},
    year = {2014},
    journal = {Nature Nanotechnology},
    author = {Muhonen, Juha T and Dehollain, Juan P and Laucht, Arne and Hudson, Fay E and Kalra, Rachpon and Sekiguchi, Takeharu and Itoh, Kohei M and Jamieson, David N and Mccallum, Jeffrey C and Dzurak, Andrew S and Morello, Andrea},
    number = {October},
    pages = {986--991},
    volume = {9},
    publisher = {Nature Publishing Group},
    doi = {10.1038/nnano.2014.211}
}

@article{Chiu2025ContinuousSystem,
    title = {{Continuous operation of a coherent 3,000-qubit system}},
    year = {2025},
    journal = {Nature},
    author = {Chiu, Neng Chun and Trapp, Elias C. and Guo, Jinen and Abobeih, Mohamed H. and Stewart, Luke M. and Hollerith, Simon and Stroganov, Pavel L. and Kalinowski, Marcin and Geim, Alexandra A. and Evered, Simon J. and Li, Sophie H. and Lyu, Xingjian and Peters, Lisa M. and Bluvstein, Dolev and Wang, Tout T. and Greiner, Markus and Vuleti{\'{c}}, Vladan and Lukin, Mikhail D.},
    number = {8087},
    month = {10},
    pages = {1075--1080},
    volume = {646},
    publisher = {Nature Research},
    doi = {10.1038/s41586-025-09596-6},
    issn = {14764687},
    pmid = {40953821},
    arxivId = {2506.20660}
}

@article{vaartjes2026maximizing,
  title={Maximizing the nondemolition nature of a quantum measurement via an adaptive readout protocol},
  author={Vaartjes, Arjen and Su, Rocky Yue and O’Neill, Laura A and Steinacker, Paul and Goenka, Gauri and van Blankenstein, Mark R and Yu, Xi and Wilhelm, Benjamin and Jakob, Alexander M and Hudson, Fay E and others},
  journal={PRX Quantum},
  volume={7},
  number={2},
  pages={020330},
  year={2026},
  publisher={APS},
  doi = {10.1103/jtn1-wzyl}
}

@article{ng2011combining,
  title={Combining dynamical decoupling with fault-tolerant quantum computation},
  author={Ng, Hui Khoon and Lidar, Daniel A and Preskill, John},
  journal={Physical Review A},
  volume={84},
  number={1},
  pages={012305},
  year={2011},
  publisher={APS},
  doi={10.1103/PhysRevA.84.012305}
}

@article{zanardi1999symmetrizing,
  title={Symmetrizing evolutions},
  author={Zanardi, Paolo},
  journal={Physics Letters A},
  volume={258},
  number={2-3},
  pages={77--82},
  year={1999},
  publisher={Elsevier},
  doi={10.1016/S0375-9601(99)00365-5}
}

@article{viola1999dynamical,
  title={Dynamical decoupling of open quantum systems},
  author={Viola, Lorenza and Knill, Emanuel and Lloyd, Seth},
  journal={Physical Review Letters},
  volume={82},
  number={12},
  pages={2417},
  year={1999},
  publisher={APS}, 
  doi={10.1103/PhysRevLett.82.2417}
}

@article{duan1999suppressing,
  title={Suppressing environmental noise in quantum computation through pulse control},
  author={Duan, Lu-Ming and Guo, Guang-Can},
  journal={Physics Letters A},
  volume={261},
  number={3-4},
  pages={139--144},
  year={1999},
  publisher={Elsevier},
  doi={10.1016/S0375-9601(99)00592-7}
}

@misc{kasatkin2026quantum,
  title={Quantum Error Correction and Dynamical Decoupling: Better Together or Apart?},
  author={Kasatkin, Victor and Morford-Oberst, Mario and Vezvaee, Arian and Lidar, Daniel A},
  journal={arXiv preprint},
  eprint={2602.19042},
  year={2026}
}

@article{Grimm2020StabilizationQubit,
    title = {{Stabilization and operation of a Kerr-cat qubit}},
    year = {2020},
    journal = {Nature},
    author = {Grimm, A. and Frattini, N. E. and Puri, S. and Mundhada, S. O. and Touzard, S. and Mirrahimi, M. and Girvin, S. M. and Shankar, S. and Devoret, M. H.},
    number = {7820},
    month = {8},
    pages = {205--209},
    volume = {584},
    publisher = {Nature Research},
    doi = {10.1038/s41586-020-2587-z},
    issn = {14764687},
    pmid = {32788737}
}

@article{hansen2023accessing,
  title={Accessing the full capabilities of filter functions: Tool for detailed noise and quantum control susceptibility analysis},
  author={Hansen, Ingvild and Seedhouse, Amanda E and Saraiva, Andre and Dzurak, Andrew S and Yang, Chih Hwan},
  journal={Physical Review A},
  volume={108},
  number={1},
  pages={012426},
  year={2023},
  publisher={APS},
  doi = {10.1103/PhysRevA.108.012426}
}

@article{Ding2025QuantumQubit,
    title = {Quantum control of an oscillator with a Kerr-cat qubit},
    year = {2025},
    journal = {Nature Communications},
    author = {Ding, Andy Z. and Brock, Benjamin L. and Eickbusch, Alec and Koottandavida, Akshay and Frattini, Nicholas E. and Corti{\~{n}}as, Rodrigo G. and Joshi, Vidul R. and de Graaf, Stijn J. and Chapman, Benjamin J. and Ganjam, Suhas and Frunzio, Luigi and Schoelkopf, Robert J. and Devoret, Michel H.},
    number = {1},
    volume = {16},
    pages = {5279},
    publisher = {Nature Research},
    doi = {10.1038/s41467-025-60352-w},
    issn = {20411723},
    pmid = {40480974},
    arxivId = {2407.10940}
}

@article{Masuda2025StabilizationRefrigerator,
    title = {Stabilization of Kerr-cat qubits with quantum circuit refrigerator},
    year = {2025},
    journal = {npj Quantum Information},
    author = {Masuda, Shumpei and Kamimura, Shunsuke and Yamamoto, Tsuyoshi and Aoki, Takaaki and Tomonaga, Akiyoshi},
    number = {1},
    month = {2},
    volume = {11},
    pages = {26},
    publisher = {Nature Research},
    doi = {10.1038/s41534-025-00974-6},
    issn = {20566387},
    arxivId = {2406.13957}
}

@article{lidar2014review,
  title={Review of Decoherence-Free Subspaces, Noiseless Subsystems, and Dynamical Decoupling},
  author={Lidar, Daniel A},
  journal={Quantum information and computation for chemistry},
  pages={295--354},
  year={2014},
  publisher={Wiley Online Library}
}

@article{hahn1950spin,
  title={Spin echoes},
  author={Hahn, Erwin L},
  journal={Physical Review},
  volume={80},
  number={4},
  pages={580},
  year={1950},
  publisher={APS},
  doi={10.1103/PhysRev.80.580}
}

@article{Mourik2017,
    title = {{Exploring quantum chaos with a single nuclear spin}},
    year = {2017},
    author = {Mourik, Vincent and Asaad, Serwan and Firgau, Hannes and Pla, Jarryd J. and Holmes, Catherine and Milburn, Gerard J. and McCallum, Jeffrey C. and Morello, Andrea},
    month = {3},
    journal = {Physical Review E},
    number = {98},
    doi = {10.1103/PhysRevE.98.042206},
    pages = {042206}
}

@article{Lim2023Fault-tolerantQudit,
    author = {Lim, Sumin and Liu, Junjie and Ardavan, Arzhang},
    title = {Fault-tolerant qubit encoding using a spin-7/2 qudit},
    journal = {Physical Review A},
    volume = {108},
    number = {6},
    pages = {062403},
    year = {2023},
    doi = {10.1103/PhysRevA.108.062403}
}

@article{Kusano2026Spin-CatArray,
    title = {{Spin-Cat Qubit with Biased Noise in an Optical Tweezer Array}},
    year = {2026},
    author = {Kusano, Toshi and Shibata, Kosuke and Yeh, Chih-Han and Saito, Keito and Nakamura, Yuma and Yokoyama, Rei and Kashimoto, Takumi and Takano, Tetsushi and Takasu, Yosuke and Takagi, Ryuji and Takahashi, Yoshiro},
    journal = {arXiv preprint},
    eprint = {2602.22883}
}

@article{McIntyre2026SpinQubits,
      title={Spin Kerr-cat qubits}, 
      author={Z. M. McIntyre and Daniel Loss},
      year={2026},
      eprint={2604.19687},
      journal = {arXiv preprint}, 
}

@article{tong2026learning,
  title={Learning error suppression strategies for dynamic quantum circuits},
  author={Tong, Christopher and Shirizly, Liran and Chen, Edward H and Wang, Derek S and Pokharel, Bibek},
  journal={arXiv preprint},
  eprint={2604.18734},
  year={2026}
}

@article{d2026proof,
  title={Proof of the Error Scaling for Universally Robust Dynamical Decoupling Sequences},
  author={D'Alessandro, Domenico and Singkanipa, Phattharaporn and Lidar, Daniel},
  journal={arXiv preprint},
  eprint={2604.25807},
  year={2026}
}

@article{genov2017arbitrarily,
  title={Arbitrarily accurate pulse sequences for robust dynamical decoupling},
  author={Genov, Genko T and Schraft, Daniel and Vitanov, Nikolay V and Halfmann, Thomas},
  journal={Physical Review Letters},
  volume={118},
  number={13},
  pages={133202},
  year={2017},
  publisher={APS},
  doi={10.1103/PhysRevLett.118.133202}
}

@article{iiyama2024qudit,
  title={Qudit-generalization of the qubit echo and its application to a qutrit-based Toffoli gate},
  author={Iiyama, Yutaro and Jang, Wonho and Kanazawa, Naoki and Sawada, Ryu and Onodera, Tamiya and Terashi, Koji},
  journal={arXiv preprint},
  year={2024},
  eprint={2405.14752}
}

@article{yuan2022preserving,
  title={Preserving multilevel quantum coherence by dynamical decoupling},
  author={Yuan, Xinxing and Li, Yue and Zhang, Mengxiang and Liu, Chang and Zhu, Mingdong and Qin, Xi and Vitanov, Nikolay V and Lin, Yiheng and Du, Jiangfeng},
  journal={Physical Review A},
  volume={106},
  number={2},
  pages={022412},
  year={2022},
  publisher={APS},
  doi={10.1103/PhysRevA.106.022412}
}

@article{zhou2024robust,
  title={Robust Hamiltonian engineering for interacting qudit systems},
  author={Zhou, Hengyun and Gao, Haoyang and Leitao, Nathaniel T and Makarova, Oksana and Cong, Iris and Douglas, Alexander M and Martin, Leigh S and Lukin, Mikhail D},
  journal={Physical Review X},
  volume={14},
  number={3},
  pages={031017},
  year={2024},
  publisher={APS},
  doi={10.1103/PhysRevX.14.031017}
}

@article{souza2011robust,
  title={Robust dynamical decoupling for quantum computing and quantum memory},
  author={Souza, Alexandre M and Alvarez, Gonzalo A and Suter, Dieter},
  journal={Physical Review Letters},
  volume={106},
  number={24},
  pages={240501},
  year={2011},
  publisher={APS},
  doi={10.1103/PhysRevLett.106.240501}
}

@article{viola2003robust,
  title={Robust dynamical decoupling of quantum systems with bounded controls},
  author={Viola, Lorenza and Knill, Emanuel},
  journal={Physical Review Letters},
  volume={90},
  number={3},
  pages={037901},
  year={2003},
  publisher={APS},
  doi={10.1103/PhysRevLett.90.037901}
}

@article{meiboom1958modified,
  title={Modified Spin-Echo Method for Measuring Nuclear Relaxation Times},
  author={Meiboom, S and Gill, D},
  journal={Review of Scientific Instruments},
  volume={29},
  number={8},
  pages={688--691},
  year={1958},
  publisher={AIP Publishing},
  doi={10.1063/1.1716296}
}

@article{suter2016colloquium,
  title={Colloquium: Protecting quantum information against environmental noise},
  author={Suter, Dieter and {\'A}lvarez, Gonzalo A},
  journal={Reviews of Modern Physics},
  volume={88},
  number={4},
  pages={041001},
  year={2016},
  publisher={APS},
  doi={10.1103/RevModPhys.88.041001}
}

@article{cywinski2008enhance,
  title={How to enhance dephasing time in superconducting qubits},
  author={Cywi{\'n}ski, {\L}ukasz and Lutchyn, Roman M and Nave, Cody P and Das Sarma, S},
  journal={Physical Review B},
  volume={77},
  number={17},
  pages={174509},
  year={2008},
  publisher={APS},
  doi={10.1103/PhysRevB.77.174509}
}

@article{bylander2011noise,
  title={Noise spectroscopy through dynamical decoupling with a superconducting flux qubit},
  author={Bylander, Jonas and Gustavsson, Simon and Yan, Fei and Yoshihara, Fumiki and Harrabi, Khalil and Fitch, George and Cory, David G and Nakamura, Yasunobu and Tsai, Jaw-Shen and Oliver, William D},
  journal={Nature Physics},
  volume={7},
  number={7},
  pages={565--570},
  year={2011},
  publisher={Nature Publishing Group UK London},
  doi={10.1038/nphys1994}
}

@article{medford2012scaling,
  title={Scaling of dynamical decoupling for spin qubits},
  author={Medford, J and Cywi{\'n}ski, {\L}ukasz and Barthel, C and Marcus, CM and Hanson, MP and Gossard, AC},
  journal={Physical Review Letters},
  volume={108},
  number={8},
  pages={086802},
  year={2012},
  publisher={APS},
  doi={10.1103/PhysRevLett.108.086802}
}

@article{biercuk2009optimized,
  title={Optimized dynamical decoupling in a model quantum memory},
  author={Biercuk, Michael J and Uys, Hermann and VanDevender, Aaron P and Shiga, Nobuyasu and Itano, Wayne M and Bollinger, John J},
  journal={Nature},
  volume={458},
  number={7241},
  pages={996--1000},
  year={2009},
  publisher={Nature Publishing Group UK London},
  doi={10.1038/nature07951}
}

@misc{keras,
  title={Keras},
  author={Chollet, Fran\c{c}ois and others},
  year={2015},
  howpublished={\url{https://keras.io}}
}

@misc{tensorflow,
title={ {TensorFlow}: Large-Scale Machine Learning on Heterogeneous Systems},
url={https://www.tensorflow.org/},
note={Software available from tensorflow.org},
author={
    Mart\'{\i}n~Abadi and
    Ashish~Agarwal and
    Paul~Barham and
    Eugene~Brevdo and
    Zhifeng~Chen and
    Craig~Citro and
    Greg~S.~Corrado and
    Andy~Davis and
    Jeffrey~Dean and
    Matthieu~Devin and
    Sanjay~Ghemawat and
    Ian~Goodfellow and
    Andrew~Harp and
    Geoffrey~Irving and
    Michael~Isard and
    Yangqing Jia and
    Rafal~Jozefowicz and
    Lukasz~Kaiser and
    Manjunath~Kudlur and
    Josh~Levenberg and
    Dandelion~Man\'{e} and
    Rajat~Monga and
    Sherry~Moore and
    Derek~Murray and
    Chris~Olah and
    Mike~Schuster and
    Jonathon~Shlens and
    Benoit~Steiner and
    Ilya~Sutskever and
    Kunal~Talwar and
    Paul~Tucker and
    Vincent~Vanhoucke and
    Vijay~Vasudevan and
    Fernanda~Vi\'{e}gas and
    Oriol~Vinyals and
    Pete~Warden and
    Martin~Wattenberg and
    Martin~Wicke and
    Yuan~Yu and
    Xiaoqiang~Zheng},
  year={2015},
}

@inproceedings{Adam,
  author    = {Diederik P. Kingma and
               Jimmy Ba},
  editor    = {Yoshua Bengio and
               Yann LeCun},
  title     = {Adam: {A} Method for Stochastic Optimization},
  booktitle = {3rd International Conference on Learning Representations, {ICLR} 2015,
               San Diego, CA, USA, May 7-9, 2015, Conference Track Proceedings},
  year      = {2015},
  url       = {http://arxiv.org/abs/1412.6980},
  bibsource = {dblp computer science bibliography, https://dblp.org}
}

@article{theory_paper,
    author = {Akram Youssry and Ritik Sareen and Andrea Morello},
    title = {Optimal control of qudits in silicon-donor platform},
    journal={},
    note = {In preparation (2026)},
}

@article{model_selection,
  title = {Predicting Non-Markovian Superconducting-Qubit Dynamics from Tomographic Reconstruction},
  author = {Zhang, Haimeng and Pokharel, Bibek and Levenson-Falk, E.M. and Lidar, Daniel},
  journal = {Phys. Rev. Appl.},
  volume = {17},
  issue = {5},
  pages = {054018},
  numpages = {22},
  year = {2022},
  month = {May},
  publisher = {American Physical Society},
  doi = {10.1103/PhysRevApplied.17.054018},
  url = {https://link.aps.org/doi/10.1103/PhysRevApplied.17.054018}
}

@article{QNS,
  title = {Measuring the Spectrum of Colored Noise by Dynamical Decoupling},
  author = {\'Alvarez, Gonzalo A. and Suter, Dieter},
  journal = {Physical Review Letters},
  volume = {107},
  issue = {23},
  pages = {230501},
  numpages = {5},
  year = {2011},
  month = {Nov},
  publisher = {American Physical Society},
  doi = {10.1103/PhysRevLett.107.230501},
  url = {https://link.aps.org/doi/10.1103/PhysRevLett.107.230501}
}

@incollection{haeberlen,
author = {Haeberlen, Ulrich},
booktitle = {Advances in Magnetic Resonance},
title = {High Resolution {NMR} in Solids Selective Averaging},
publisher = {Academic Press},
year = {1976},
isbn = {978-0-12-025561-0},
doi = {https://doi.org/10.1016/B978-0-12-025561-0.50001-0},
url = {https://www.sciencedirect.com/science/article/pii/B9780120255610500010}
}

@article{wang2020qudits,
  title={Qudits and high-dimensional quantum computing},
  author={Wang, Yuchen and Hu, Zixuan and Sanders, Barry C and Kais, Sabre},
  journal={Frontiers in Physics},
  volume={8},
  pages={589504},
  year={2020},
  publisher={Frontiers},
  doi = {10.3389/fphy.2020.589504}
}

@article{yu2025schrodinger,
  title={Schr{\"o}dinger cat states of a nuclear spin qudit in silicon},
  author={Yu, Xi and Wilhelm, Benjamin and Holmes, Danielle and Vaartjes, Arjen and Schwienbacher, Daniel and Nurizzo, Martin and Kringh{\o}j, Anders and Blankenstein, Mark R van and Jakob, Alexander M and Gupta, Pragati and others},
  journal={Nature Physics},
  volume={21},
  number={3},
  pages={362--367},
  year={2025},
  publisher={Nature Publishing Group UK London},
  doi = {10.1038/s41567-024-02745-0}
}

@article{du2009preserving,
  title={Preserving electron spin coherence in solids by optimal dynamical decoupling},
  author={Du, Jiangfeng and Rong, Xing and Zhao, Nan and Wang, Ya and Yang, Jiahui and Liu, RB},
  journal={Nature},
  volume={461},
  number={7268},
  pages={1265--1268},
  year={2009},
  publisher={Nature Publishing Group UK London},
  doi = {10.1038/nature08470}
}

\end{document}


\title{Supplementary Information: Dynamical Decoupling of Nonclassical States in a Biased-Noise Spin Qudit}

\date{\today}

\maketitle

\tableofcontents

\newcommand{\beginsupplement}{%
        \setcounter{table}{0}
        \renewcommand{\thetable}{S\arabic{table}}%
        \setcounter{figure}{0}
        \renewcommand{\thefigure}{S\arabic{figure}}%
        \setcounter{section}{0}
        \renewcommand*{\thesection}{S\arabic{section}}%
        \setcounter{equation}{0}
        \renewcommand{\theequation}{S\arabic{equation}}%
     }
\newcommand{\Sb}{^{123}\rm Sb}
\beginsupplement
\section{Definitions of the $d=8$ spin operators}
\label{sec:spin_operators}

Here, we provide an overview of the spin operators of a spin-$I=7/2$ $\Sb$ system ($d=2I+1=8$). The operators are defined in the
$\{\ket{m_I}\}$ basis ordered as
$\bigl\{\ket{7/2},\ket{5/2},\ket{3/2},\ket{1/2},
\ket{-1/2},\ket{-3/2},\ket{-5/2},\ket{-7/2}\bigr\}$.

\begin{equation}
\hat{I}_x =
\frac{1}{2}
\begin{pmatrix}
0 & \sqrt{7} & 0 & 0 & 0 & 0 & 0 & 0 \\
\sqrt{7} & 0 & \sqrt{12} & 0 & 0 & 0 & 0 & 0 \\
0 & \sqrt{12} & 0 & \sqrt{15} & 0 & 0 & 0 & 0 \\
0 & 0 & \sqrt{15} & 0 & 4 & 0 & 0 & 0 \\
0 & 0 & 0 & 4 & 0 & \sqrt{15} & 0 & 0 \\
0 & 0 & 0 & 0 & \sqrt{15} & 0 & \sqrt{12} & 0 \\
0 & 0 & 0 & 0 & 0 & \sqrt{12} & 0 & \sqrt{7} \\
0 & 0 & 0 & 0 & 0 & 0 & \sqrt{7} & 0
\end{pmatrix}.
\label{appendix_eq:Ix}
\end{equation}

\begin{equation}
\hat{I}_y =
\frac{1}{2i}
\begin{pmatrix}
0 & \sqrt{7} & 0 & 0 & 0 & 0 & 0 & 0 \\
-\sqrt{7} & 0 & \sqrt{12} & 0 & 0 & 0 & 0 & 0 \\
0 & -\sqrt{12} & 0 & \sqrt{15} & 0 & 0 & 0 & 0 \\
0 & 0 & -\sqrt{15} & 0 & 4 & 0 & 0 & 0 \\
0 & 0 & 0 & -4 & 0 & \sqrt{15} & 0 & 0 \\
0 & 0 & 0 & 0 & -\sqrt{15} & 0 & \sqrt{12} & 0 \\
0 & 0 & 0 & 0 & 0 & -\sqrt{12} & 0 & \sqrt{7} \\
0 & 0 & 0 & 0 & 0 & 0 & -\sqrt{7} & 0
\end{pmatrix}.
\end{equation}

\begin{equation}
\hat{I}_z =
\begin{pmatrix}
\frac{7}{2} & 0 & 0 & 0 & 0 & 0 & 0 & 0 \\
0 & \frac{5}{2} & 0 & 0 & 0 & 0 & 0 & 0 \\
0 & 0 & \frac{3}{2} & 0 & 0 & 0 & 0 & 0 \\
0 & 0 & 0 & \frac{1}{2} & 0 & 0 & 0 & 0 \\
0 & 0 & 0 & 0 & -\frac{1}{2} & 0 & 0 & 0 \\
0 & 0 & 0 & 0 & 0 & -\frac{3}{2} & 0 & 0 \\
0 & 0 & 0 & 0 & 0 & 0 & -\frac{5}{2} & 0 \\
0 & 0 & 0 & 0 & 0 & 0 & 0 & -\frac{7}{2}
\end{pmatrix}.
\end{equation}

\section{Qudit DD theory}

\subsection{Qubit DD and Hahn echo derivation}

Let us remind ourselves how one of the simplest DD sequences, the ``XIXI'' Hahn echo DD sequence works to suppress dephasing noise in a single qubit system \cite{lidar_lecture_2020}. We consider a simple system-bath Hamiltonian of the form,
\begin{align}
H_{\text{SB}} = \sigma^{z} \otimes B^{z},
\end{align}
where \(B^{z}\) is some bath operator and we have suppressed the system and bath Hamiltonians for ease. Let \(f_{\tau} := e^{-i H_{\text{SB}}\tau}\) denote the ``free evolution'' of the qubit under the system-bath coupling. Then, we see a ``bang-bang'' pulse of the form,
\begin{align}
X f_{\tau} X f_{\tau} = I.
\end{align}
For the Hamiltonain above, the cancellation is exact for ideal instantaneous pulses. To do this, we just need to remind ourselves of two simple identities: the anti-commutation of the Pauli's and the fact that exponential functions are unitary automorphisms \(U e^{A} U^{\dagger} = e^{U A U^{\dagger}}\). For a simple system-bath Hamiltonian above, one can show that choosing \(X=\sigma_x\) leads to 
\begin{align*}
\sigma^{x} f_{\tau} \sigma^{x} f_{\tau} &= \exp \left[-i \tau \left( \sigma^{x} \sigma^{z} \sigma^{x} \right) \otimes B^{z} \right] f_{\tau} \\
&= \exp \left[ + i \tau \sigma^{z} \otimes B^{z} \right] f_{\tau} \\
&= f_{-\tau} f_{\tau} = I.
\end{align*}

\subsection{Qudit generalization of the Hahn echo}
\label{sec:global_hahn_echo}

With this, we now generalize the Hahn echo to the qudit case. Our system of interest is a spin-\(7/2\) nuclear spin qudit for which the primary noise is dephasing \cite{yu2025schrodinger,Fernandez2024, Mourik2017,Savytskyy2023}. So we can write down a simple system-bath model of the form,
\begin{align}
H_{\text{SB}} = h I_{z} \otimes B + q I^{2}_{z} \otimes \tilde{B},
\end{align}
where \(I_{\alpha}, \alpha = \{ x,y,z \}\) are the spin angular momentum operators defined in~\ref{sec:spin_operators} and \(B, \tilde{B}\) are not necessarily commuting bath operators. In our case we are looking for a \(d = 2I+1\) dimensional representation, therefore our qudit dimension is \(8\). The \(h\) coupling is called the ``magnetic'' term (or the Zeeman noise) and the \(q\) coupling is called the ``quadrupole'' term (or the electric noise).

We will primarily focus on a pulse sequence to cancel out the first term. Just like the single qubit case, if we can find an operator \(X_{s}\) such that
\begin{align}
\{ X_{s}, I_{z} \} = 0 \iff X_{s} I_{z} X^{\dagger}_{s} = - I_{z}.
\end{align}
Here $X_s$ is unitary, Hermiticity is not required.

An interesting thing about the unitary \(X_{s}\) above is that it \textit{does not} affect the quadrupole term at all since,
\begin{align}
X_{s} I^{2}_{z} X^{\dagger}_{s} = X_{s} I_{z} \underbrace{X^{\dagger}_{s} X_{s}}_{= I} I_{z} X^{\dagger}_{s} = I^{2}_{z}.
\end{align}
So we can cancel the magnetic term without changing the quadrupole term at all, which could be employed in a "concatenated DD" sequence to first cancel \(I_z\) noise and then the \(I_z^2\) noise.

One can show that the \(I_z\) noise is canceled out by a \(SU(2)\) rotation along the \(x\)-axis, namely,
\begin{align}
    X_s = R_x(\pi)=e^{i \pi I_x}.
\end{align}
That is, a simple Hahn echo sequence cancels the \(I_z\) noise, which generalizes the qubit case of canceling \(\sigma_{z}\) dephasing with pulse along \(\sigma_{x}\).

\subsection{Clock and Shift unitaries}

The clock-and-shift matrices for a single qudit system, \(\mathcal{H} \cong \mathbb{C}^{d}\) are the generalizations of the standard qubit Pauli \(\sigma_{x}\) and \(\sigma_{z}\).
\begin{align}
Z_{d} | j \rangle &= \omega^{j} | j \rangle, \quad \omega = e^{\frac{2 \pi i}{d}},\\
X_{d} | j \rangle &= | j+1 \rangle \quad  \mod d, \\
Z_d X_d &=\omega X_d Z_d
\end{align}
Note that the generator matrix \(X_d\) in the main text is written in the descending matrix notation (it is the transpose of the matrix here).

Together, the qudit Paulis generate the Heisenberg-Weyl group for a single qudit,
\begin{align}
\mathcal{H} \mathcal{W}_d=\left\{\omega^c X_d^a Z_d^b: a, b, c \in \mathbb{Z}_d\right\}, \quad \omega=e^{2 \pi i / d}.
\end{align}
The generators additionally satisfy the property \(\left( X_{d} \right)^{d} = \mathbb{I} = \left( Z_{d} \right)^{d}\). We use the notation with a subscript \(d\) to emphasize that they are a \textit{qudit} Pauli and not the standard ones (these are not Hermitian, only unitary).

If the system-bath coupling above instead had the form, \(H_{\mathrm{SB}}=h Z_d \otimes B+h^* Z_d^{\dagger} \otimes B^{\dagger}\) (i.e., \(I_z \mapsto Z_{d}\)) then we have the following DD sequence that anticommutes with \(Z_{d}\). We first note the following lemma \cite{nielsen_universal_2002},
\begin{align}
\left( X^{j}_{d} Z^{k}_{d} \right) \left( X^{s}_{d} Z^{t}_{d} \right) = \omega^{ks-jt} \left( X^{s}_{d} Z^{t}_{d} \right) \left( X^{j}_{d} Z^{k}_{d} \right).
\end{align}
Setting \(j=0=t\) gives us,
\begin{align}
\left( X^{s}_{d} \right)^{\dagger} Z^{k}_{d} X^{s}_{d} = \omega^{ks} Z^{k}_{d}.
\end{align}
This is the identity we need. For even \(d\), setting \(k=1\) and \(s = d/2\) gives us the anti-commutation relation,
\begin{align}
\left( X^{d/2}_{d} \right)^{\dagger} Z_{d} X^{d/2}_{d} = - Z_{d}.
\end{align}
Therefore, we note that for a qudit, the Pauli \(X_{d}^{d/2}\) anti-commutes with the generalized \(Z_{d}\). Setting \(d=2\) gives us the qubit result back. Note that, this inversion does not reverse even powers of $Z_d$ and therefore does not cancel arbitrary diagonal noise.

\subsection{Decoupling groups and 1-designs}

If we want to decouple an arbitrary system-bath interaction, then we need to average over a unitary 1-design \cite{anand2026quditdesigns}. These can be equivalently thought of as a set of unitaries that form a complete orthonormal basis for the operator space \(\mathcal{L}(\mathcal{H})\) such as the Pauli group (whilst being careful of phases etc). 

We assume a \textit{traceless} Hamiltonians, i.e., since we want \(H \mapsto -H\) under some (unitary) pulse sequence, the trace of the Hamiltonian must be zero. For qudits, one can make the Hamiltonian traceless by subtracting the trace part as
\begin{align}
\label{eq:traceless-hamiltonian}
H \mapsto H - \operatorname{Tr}\left[ H \right] \frac{\mathbb{I}}{d},
\end{align}
which is equivalent to shifting the eigenvalues as \(\lambda_{j} \mapsto \lambda_{j} - \left( \sum_{j} \lambda_{j} \right)/d\). That is, all of the energies of the Hamiltonian are shifted by the ``mean energy''. 

The ``group averaging'' of the traceless Hamiltonian gives us,
\begin{align*}
U_{\mathrm{eff}} = \prod_{g \in G} g e^{-iHt} g^{\dagger} = \prod_{g \in G} \exp \left[ -i g H g^{\dagger} t \right]  \approx \exp \left[ -it \sum\limits_{g \in G} g H g^{\dagger} \right] = \mathbb{I},
\end{align*}
where we have made a first-order Trotter approximation. Since the elements \(g \in G\) form a one-design, averaging over them is equivalent to \(H \mapsto \operatorname{Tr}\left[ H \right] \mathbb{I} / d = 0\); assuming the system Hamiltonian is traceless, and so the system evolution is trivialized. Applied to a system-bath Hamiltonian, the average is \(\mathbb{I}_d \otimes \operatorname{Tr}_s(H) / d\). For the first-order Trotter approximation to hold, the DD pulses have to be much faster than the system-bath interaction (a common assumption in DD theory). 

What is also useful here is that the minimum number of elements of a discrete, exact, unitary 1-design are exactly equal to \(d^2 = 4^n\) (for \(n\) qubits), which is exponential in the number of qubits. This further exemplifies the need for noise-adapted dynamical decoupling sequences. 

Also notice that since the qudit Pauli group forms a unitary \(1\)-design in every dimension \(d \in \mathbb{N}\), this scheme always works (although one should be careful in using this for CV systems) \cite{anand2026quditdesigns}. Moreover, even if the Hamiltonian above is not traceless, the effect of the 1-design DD sequence is to \(H \mapsto \operatorname{Tr}\left[ H \right] \mathbb{I}/d\), which gives us an identity operator on the system part, up to a phase factor, which can be absorbed into a bath term. It is in this sense that 1-design DD works to decouple an arbitrary system-bath Hamiltonian.

\subsection{Noise-adapted DD and weighted spherical designs}
\label{sec:dd_weighted_spherical_designs}

For a fixed system--bath Hamiltonian and a closed cycle of ideal instantaneous
pulses, let $g_k$ denote the cumulative control during the $k$th free-evolution
interval. With $T=\sum_k\tau_k$ and $w_k=\tau_k/T>0$, the leading Magnus term
averages each system operator according to
\begin{equation}
    \Pi(A)=\sum_{k=1}^{N}w_k g_k^\dagger A g_k,
    \qquad \sum_{k=1}^{N}w_k=1.
    \label{eq:dd_weighted_average}
\end{equation}
Decoupling requires $\Pi(A)=\operatorname{Tr}(A)I_d/d$ only for operators in
the noise space; a unitary $1$-design imposes this condition on the entire
operator space. This distinction underlies the group-factorization approach
of Read \emph{et al.}~\cite{read2025groupfactorization}.

The weighted-design viewpoint of Ref.~\cite{anand2026quditdesigns} gives a
useful extension beyond group averages. Consider the axial noise space
$\mathcal E_r=\operatorname{span}\{I_d,I_z,\ldots,I_z^r\}$, where
$1\leq r\leq 2I$ is an integer and $d=2I+1$. For spin rotations $g_k$, define
unit vectors $\mathbf n_k$ by
$g_k^\dagger I_z g_k=\mathbf n_k\cdot\mathbf I$, with
$\mathbf I=(I_x,I_y,I_z)$. Then $\Pi$ averages every operator in
$\mathcal E_r$ to a scalar if and only if the weighted directions form a
spherical $r$-design, i.e.,
\begin{equation}
    \sum_{k=1}^{N}w_kY_{\ell m}(\mathbf n_k)=0,
    \qquad 1\leq\ell\leq r,\quad -\ell\leq m\leq\ell,
    \label{eq:dd_spherical_design}
\end{equation}
where $Y_{\ell m}$ are spherical harmonics. To see this, the axial irreducible
tensor operators $T_{\ell0}$, $0\leq\ell\leq r$, form a basis of
$\mathcal E_r$: each is a polynomial of degree $\ell$ in $I_z$ with nonzero
leading coefficient. Its rotated tensor components span the degree-$\ell$
spherical harmonics. All nonscalar
components therefore vanish precisely when
Eq.~\eqref{eq:dd_spherical_design} holds. The weights have the direct meaning
of dwell fractions, and the rotations need not form a group.

For the magnetic and quadrupolar noise considered here ($r=2$), the design
conditions reduce to
\begin{equation}
    \sum_k w_k\mathbf n_k=0,
    \qquad
    \sum_k w_k\mathbf n_k\mathbf n_k^{\mathsf T}=\frac{I_3}{3}.
    \label{eq:dd_spherical_second_moment}
\end{equation}
They imply $\Pi(I_z)=0$ and
$\Pi(I_z^2)=I(I+1)I_d/3$, equal to $21I_8/4$ for $I=7/2$.
Thus $H_{SB}=hI_z\otimes B+qI_z^2\otimes\widetilde B$ averages to
$qI(I+1)I_d\otimes\widetilde B/3$, which acts only on the bath.
Four equally weighted tetrahedral directions suffice. They can be generated
by the four cumulative rotations
\begin{equation}
    \{I_d,e^{-i\pi I_{\mathbf a}},e^{-i\pi I_{\mathbf b}},
    e^{-i\pi I_{\mathbf c}}\},
    \qquad
    \hat{\mathbf z}=\frac{\mathbf a+\mathbf b+\mathbf c}{\sqrt{3}},
    \label{eq:dd_four_rotations}
\end{equation}
where $\mathbf a,\mathbf b,\mathbf c$ are orthonormal and
$I_{\mathbf a}=\mathbf a\cdot\mathbf I$, and similarly for
$\mathbf b,\mathbf c$. This is the $\mathrm D_2$ construction of
Ref.~\cite{read2025groupfactorization}, whose closed four-pulse cycle
alternates $\pi$ rotations about $\mathbf a$ and $\mathbf b$ with equal
free-evolution intervals. Closure is up to a global phase, which does not
affect Eq.~\eqref{eq:dd_weighted_average}.
Four intervals are minimal within this spin-rotation setting for $I\geq1$:
the second moment has rank three, while the zero first moment bounds this
rank by $N-1$. Nevertheless, $I_{\mathbf a}^2$ is invariant and nonscalar,
so these rotations do not form a qudit unitary $1$-design. This construction
suppresses the modeled noise at leading Magnus order; the cyclic DD sequence
used here has the broader property of averaging every diagonal operator to
a scalar.

\subsection{Reduction in DD overhead for biased-noise qudits}
It is worth clarifying the nature of reduction in dynamical decoupling resources for biased-noise qudits (from \(d^2\) to \(d\) in our case). There are two different quantities of interest when trying to quantify this reduction: the order of the decoupling group and the total number of elementary gates/pulses needed for DD. As we will argue, dephasing noise enables a quadratic reduction (from \(d^2\) to \(d\)) for both quantities.

Given a specific system-bath interaction, one can uniquely identify a minimal group algebra that decouples the system from the bath. The underlying group here is called the decoupling group~\cite{zanardi1999symmetrizing,lidar_lecture_2020}. As an example, for an arbitrary system-bath coupling, one can show that the decoupling group must contain at least \(d^2\) unitaries, such as those originating from a unitary 1-design~\cite{anand2026quditdesigns}. For this discussion, we focus on the projective group (modding out global phases), otherwise the order of the group is \(d^3\); the extra factor of \(d\) originates from \(d\) phases needed to turn the set of unitaries into a group (closure under multiplication) but can be ignored for this discussion.

Similarly, for arbitrary dephasing noise, where the group algebra has precisely \(d\) generators (there are \(d\) independent diagonal entries), the decoupling group is also abelian (since the noise is abelian) and has order \(d\). The fact that the structure of the noise informs the structure of the decoupling group is a deep fact that originates from thinking about DD as a symmetrization principle~\cite{zanardi1999symmetrizing}.

Note that a discussion of the decoupling group does not immediately translate to the quantum resources required to implement a DD protocol in the lab. A typical element of the decoupling group needs to be further decomposed into some elementary gate set for the qudit, such as SNAPs and displacements or SNAPs and \(SU(2)\) pulses (two-level NMR pulses)~\cite{PhysRevA.71.052318,PhysRevA.92.040303,yu2025schrodinger}. When discussing the reduction in the overhead for decoupling biased-noise qudits, it is important to clarify whether one is comparing the order of the decoupling group or the number of elementary gates. We obtain a reduction by a factor of \(d\), independent of which quantifier we use.

For example, using SNAP-\(SU(2)\)'s every qudit unitary can be decomposed uniquely in terms of \(O(d^2)\) elementary gates. If the system-bath interaction is arbitrary then we need \(d^2\) decoupling group (global) unitaries, each of which will have a decomposition with \(O(d^2)\) elementary pulses. Then the total number of elementary pulses needed to cancel arbitrary system-bath coupling is \(O(d^4)\). However, for biased-noise qudits, the decoupling group has order \(d\) and hence the total number of elementary pulses is \(O(d^3)\). Therefore, we obtain a reduction of quantum resouces by a factor of \(d\), independent of whether we count the order of the decoupling group or the number of elementary gates/pulses.

\section{Nuclear spin readout and initialization}

We read out the nuclear spin state via an adaptive readout protocol that is discussed in detail in Ref.~\cite{vaartjes2026maximizing}. 

\begin{figure}
    \centering
    \includegraphics[width=0.8\linewidth]{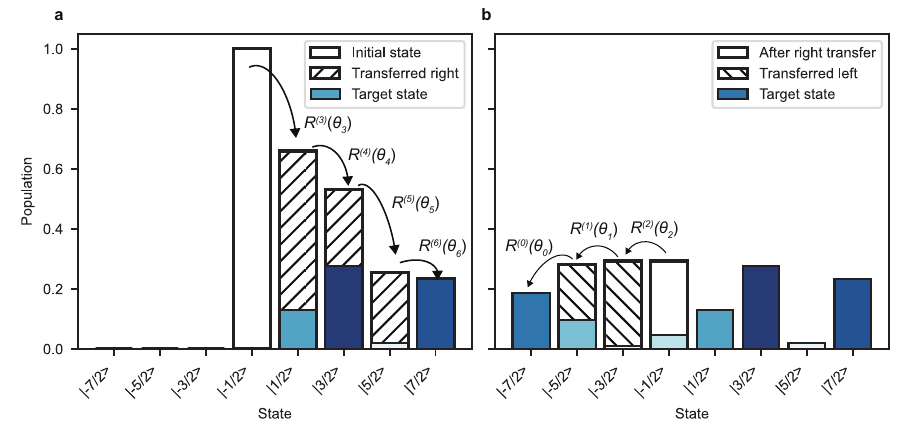}
    \caption{\textbf{Arbitrary state initialization with ladder operations. a)} We start from an arbitrary initial eigenstate, here $\ket{-1/2}$ is chosen. We first transfer the population that needs to be distributed to higher spin states to the right in this graph using sequential two-level ladder operations $R^{(k)}$ (see section~\ref{methods_sec:givens_rotations}). After the population is distributed correctly, we achieve the desired phases with a single virtual-SNAP gate.}
    \label{fig:arbitrary_initialization}
\end{figure}

We initialize any arbitrary spin state using a combination between two-level Givens rotations (section~\ref{methods_sec:givens_rotations}) and virtual-SNAP gates (section~\ref{methods_sec:vsnap_gates}). Figure~\ref{fig:arbitrary_initialization} shows a graphic representation of the initialization algorithm. To initialize an arbitrary state $\ket{\psi} =\sum_{m=-7/2}^{7/2} c_{m} \ket{m_I}$, we execute the following 4 subroutines:

\begin{enumerate}
    \item Measure the nuclear spin state $m_I=m_{\rm meas}$. In the example (Fig.~\ref{fig:arbitrary_initialization}), this has resulted in the state $\ket{-1/2}$ being populated.
    \item Calculate the necessary population transfer to the higher spin states $\ket{m_I}$, with $m_I>m_{\rm meas}$. Transfer the population with sequential givens rotations around the $-y$-axis, denoted $R_{-y}^{(k)}(\theta_k)$, and leave behind the desired population in each eigenstate. 
    \item Then, redistribute the remaining population in $m_I=m_{\rm meas}$ to the lower spin states $\ket{m_I}$, with $m_I<m_{\rm meas}$. 
    \item Finally, because all the rotations were performed about the $-y$-axis, all relative phases $\varphi_m = \arctan({\rm Re}(c_m))/{\rm Im}(c_m)=0$. Perform a virtual-SNAP gate by updating the GRF (see section~\ref{methods_sec:vsnap_gates}) to reach the desired phases.
\end{enumerate}

\section{Qudit control}

\subsection{Generalized Rotating Frame}
\label{methods_sec:GRF}

The DD experiments are described in a Generalized Rotating Frame. Our starting point for defining the GRF is the static spin Hamiltonian:
\begin{equation}
\label{eq:Hamiltonian}
\hat{{H}}^+ = -\gamma_{\rm n}B_{0}\hat{I}_z+\sum_{\alpha,\beta\in\{x,y,z\}} Q^+_\mathrm{\alpha\beta} \hat{I}_\mathrm{\alpha}\hat{I}_\mathrm{\beta},
\end{equation}
where $B_0 =$ \SI{1.384}{\tesla} is a static magnetic field, $\gamma_{\rm n}= \SI{5.55}{\mega \hertz/\tesla}$ is the nuclear gyromagnetic ratio, $\hat{I}_{\{x, y, z\}}$ are the $D=8$ nuclear spin operators and $Q^+_\mathrm{\alpha\beta}$ are the components of the $3\times3$ quadrupole tensor.
The interaction between the 8 spin eigenstates is generated by Nuclear Magnetic Resonance (NMR) and is described by the following interaction Hamiltonian:
\begin{equation}
  \hat{H}_\mathrm{1}(t) = -\gamma_\mathrm{n} \hat{I}_x \sum_{k=0}^{2I-1}\cos(2\pi f_k t + \phi_k)B_{1,k}(t),
  \label{eq:H_1}
\end{equation}
where $f_k$ are the NMR frequencies, $B_{1,k}(t)$ the oscillating magnetic field amplitudes, and $\phi_k$ the physical drive phases. The GRF \cite{Leuenberger2003,godfrinOperatingQuantumStates2017, yu2025schrodinger} 
which changes the basis to the rotating eigenstates $\exp(-i 2 \pi E_{m} t / \hbar)\ket{m_I}$, where $E_{m}$ are the eigenenergies of the static Hamiltonian \cite{Neeley2009}. In this basis:
\begin{equation}
\hat{H}_{\mathrm{GRF}} = 
  \left(
  \begin{array}{ccccccc}
  0 & g_0 e^{i\phi_0} & 0 & \cdots & 0 & 0 \\
  g_0^* e^{-i\phi_0} & 0 & g_1 e^{i \phi_1} & \cdots & 0 & 0 \\
  0 & g_1^* e^{-i \phi_1} & 0 & \cdots & 0 & 0 \\
  \vdots & \vdots & \vdots & \ddots & \vdots & \vdots \\
  0 & 0 & 0 & \cdots & 0 & g_6 e^{i\phi_6} \\
  0 & 0 & 0 & \cdots & g_6^* e^{-i\phi_6} & 0 \\
  \end{array}
  \right)
\end{equation}
where $g_k=-\frac{\gamma_n}{4} B_{1,k}(t)\bra{7/2-k}\hat{I}_x\ket{7/2-(k+1)}$.

\subsection{Givens Rotations}
\label{methods_sec:givens_rotations}

In the GRF, two-level transitions - or Givens rotations - are performed by activating one $B_{1, k}$ at a time. For example, we can induce the NMR Rabi transition $\ket{7/2}\leftrightarrow\ket{5/2}$ by activating the oscillating magnetic field $B_{1,0}$ on resonance with the transition. In that case, the rotating frame Hamiltonian becomes: 

\begin{equation}
\hat{H}_{\mathrm{GRF}} = -\frac{\gamma_n}{4}
  \left(
  \begin{array}{ccccccc}
  0 & \sqrt{7}B_{1,0} e^{i\phi_0} & 0&\cdots & 0 & 0 \\
  \sqrt{7} B_{1,0} e^{-i\phi_0} & 0 & 0 & \cdots & 0 & 0 \\
  0 & 0 & 0 & \cdots & 0 & 0 \\
  \vdots & \vdots & \vdots & \ddots & \vdots & \vdots \\
  0 & 0 & 0 & \cdots & 0 & 0 \\
  0 & 0 & 0 & \cdots & 0 & 0 \\
  \end{array}
  \right).
\end{equation}

Exponentiating this Hamiltonian provides the unitary evolution of the system:

\begin{equation}
e^{-i t \hat{H}_{\mathrm{GRF}}} =
\begin{pmatrix}
\cos(\Omega_0 t/2) & -i e^{i\phi_0}\sin(\Omega_0 t/2) & 0 & \cdots & 0 & 0 \\
-i e^{-i\phi_0}\sin(\Omega_0 t/2) & \cos(\Omega_0 t/2) & 0 & \cdots & 0 & 0 \\
0 & 0 & 1 & \cdots & 0 & 0 \\
\vdots & \vdots & \vdots & \ddots & \vdots & \vdots \\
0 & 0 & 0 & \cdots & 1 & 0 \\
0 & 0 & 0 & \cdots & 0 & 1
\end{pmatrix} 
=\hat{R}_{\phi_0}^{(0)}(\theta_0).
\end{equation}

So, in the GRF, activating $B_{1,0}$ leads to a rotation between $\ket{7/2}$ and $\ket{5/2}$ with a Rabi frequency of $\Omega_0 = \sqrt{7}\,B_{1,0}\gamma_n/2$. Colloquially, these two-level unitaries, particularly $\pi$-rotations of this kind, are called ``ladder operations'', because they can be sequentially applied to climb the ladder of spin eigenstates. We denote the $k^{\rm th}$ ladder operation with $\hat{R}^{(k)}_{\phi_k}(\theta_k)$, where $\theta_k=\Omega_kt$. If the rotation is around a Cartesian axis, say the $x, y$ or $z-$axis of the Bloch sphere, we also denote this rotation by $\hat{R}^{(k)}_{\{x, y, z\}}(\theta_k)$. 

\subsection{SU(2)-rotations}
\label{methods_sec:su2_rotations}

By applying a multi-frequency drive on resonance with all 7 NMR frequencies, we achieve an SU(2)-covariant ``global rotation'' - a rigid rotation in the entire 8-dimensional spin manifold \cite{Neeley2009,yu2025schrodinger}. Applying an equal amplitude $B_{1,k}=B_1$ on all 7 tones, all with the same initial phase $\phi_k=\varphi$ leads to a SU(2)-covariant rotation. In that case, the GRF Hamiltonian can be expressed in the spin operators $\hat{I}_x$ and $\hat{I}_y$:

\begin{equation}
    \hat{H}_{\mathrm{GRF}} = -\frac{\gamma_n}{2}B_1 (\cos \varphi \hat{I}_x + \sin \varphi \hat{I}_y).
\end{equation}

The unitary evolution of this Hamiltonian describes a rotation around a $\varphi-$rotated axis on the equator of the 8-dimensional Bloch sphere:

\begin{equation}
    \hat{R}_\varphi(\theta) = e^{-i\theta(\cos(\varphi)\hat{I}_x + \sin(\varphi)\hat{I}_y)},
\end{equation}

where $\theta(t)=\frac{\gamma_n B_1 t}{2}$. For example, if $\varphi=0$ and the pulse time $t$ is chosen such that $\theta=\pi$, the unitary rotation is: 

\begin{equation}
    \hat{R}_x(\pi) = i
\begin{pmatrix}
0 & 0 & 0 & \cdots & 0 & 1 \\
0 & 0 & 0 & \cdots & 1 & 0 \\
0 & 0 & 0 & \cdots & 0 & 0 \\
\vdots & \vdots & \vdots & \ddots & \vdots & \vdots \\
0 & 1 & 0 & \cdots & 0 & 0 \\
1 & 0 & 0 & \cdots & 0 & 0
\end{pmatrix}.
\end{equation}

\begin{figure}
    \centering
    \includegraphics[width=\linewidth]{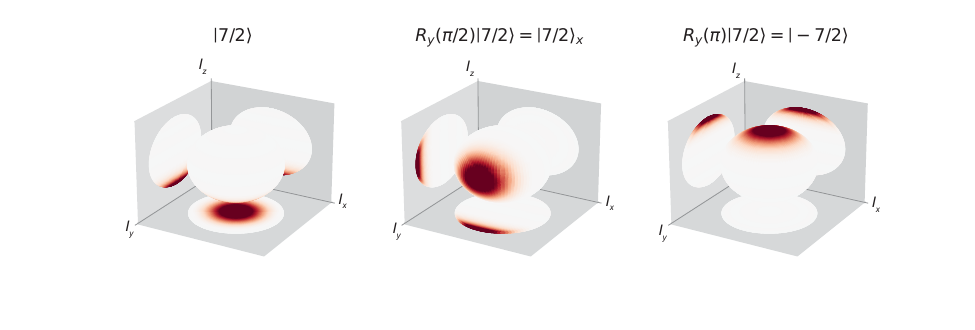}
    \caption[SU(2) covariant rotation]{\textbf{SU(2) rotation.} \textbf{a)} A multi-frequency drive consisting of all 7 NMR frequencies with equal amplitudes activates a global SU(2) covariant rotation. \textbf{a)} An initial spin-coherent state $\ket{-7/2}$ \textbf{b)} makes a rigid rotation of $\theta=\pi/2$ around the $\hat{I}_x$-axis to the equator of the Bloch sphere. The final state is $\ket{-7/2}_{y}$, i.e. a spin coherent state along $-\hat{I}_y$. 
    \textbf{c)} A rotation of $\theta=\pi$ results in the full inversion of the state. Note that throughout the whole rotation, the shape of the Wigner function stays intact.}
    \label{methods_fig:su2_rotation}
\end{figure}

This unitary constitutes an SU(2)-covariant $\pi$-pulse over the $\hat{I}_x-$axis of the 8-dimensional Bloch sphere. When applied to the spin-coherent state $\ket{-7/2}$, the resulting state will be $\ket{7/2}$, which illustrates that this $\pi$-rotation effectively inverts the 8-dimensional state space. The resulting state $\ket{7/2}$ is also a spin coherent state, indicating that the SU(2) rotations rigidly rotate the spin without changing the shape of the Wigner function (see Fig.\ref{methods_fig:su2_rotation}).

\subsection{Virtual SNAP gates}
\label{methods_sec:vsnap_gates}

In $\Sb$ nuclear qudits, selective number-dependent arbitrary phase (SNAP) gates \cite{Heeres2015CavityGates} apply distinct phases $\varphi_m$ to each eigenstate $\ket{m_I}$ of $\hat{I}_z$. They can be generally expressed as:
\begin{equation}
    \hat{U}_{\rm{SNAP}}
     \left( \Vec{\varphi} \right) = \prod_{m=-7/2}^{+7/2} e^{i\varphi_m \ket{m_I}\bra{m_I}}.
\end{equation}

\noindent Shifting the phase of the driving tones by $\Delta \phi_k$ results in a frame rotation in the GRF. The phase gate angles $\varphi_m$ relate to the frame rotation angles $\Delta \phi$ as a cumulative sum: 
\begin{equation}
    \varphi_m = - \sum_{i=0}^k{\Delta \phi_i}
    \label{phase_update}
\end{equation}

\begin{figure}
    \centering
    \includegraphics[width=0.85\linewidth]{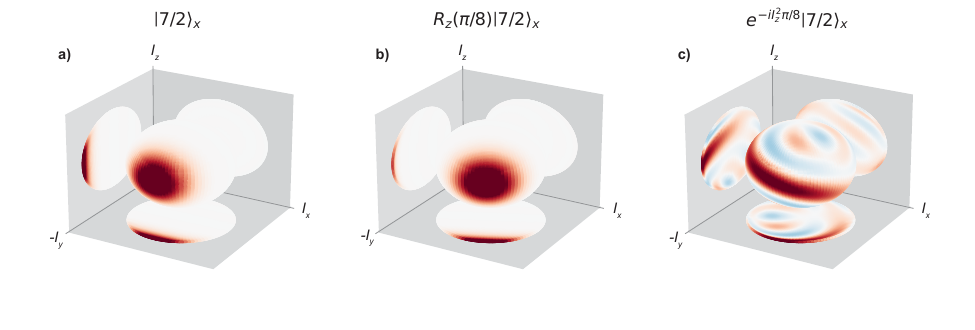}
    \caption[Virtual SNAP gates]{\textbf{Virtual SNAP gates.} An initial spin-coherent state $\ket{7/2}_x$ \textbf{b)} rotates in the GRF by virtually applying a SNAP gate. We create rigid $\hat{R}_z$ rotations by applying 7 equal phase offsets to the 7 clocks keeping track of the GRF.
    \textbf{c)} Alternatively, non-linear rotations generated by powers of $\hat{I}_z$ can be created virtually. Here we show an $\hat{I}_z^2$ rotation of $\theta=\pi/8$, creating a one-axis twisting dynamic and squeezing of the Wigner function.}
    \label{methods_fig:vsnap}
\end{figure}

\section{Noise Modelling and Simulation}
To model the Ramsey experiments of the $\Sb$ qudit shown in Fig.~1c-d of the main text, we simulate the free decay dynamics using filter function formalism \cite{theory_paper}. In particular, we start with the $d=8$ qudit Hamiltonian in the generalized rotating frame 
\begin{align}
    H(t) = \beta_z(t)I_z + \beta_{zz}(t)I_z^2
\end{align}
where we assume the classical noise processes $\beta_z(t)$ and $\beta_{zz}(t)$ are stationary, colored, and Gaussian with two-sided power spectral densities $S_z(\omega)$ and $S_{zz}(\omega)$ respectively. Under these assumptions, the coherence of an initially-prepared cat state
\begin{align}
    \ket{\psi(0)}=\frac{1}{\sqrt{2}}\left(\ket{m}+\ket{-m}\right)
\end{align}
undergoing free decay at a time $t=\tau$ is given by \cite{theory_paper}:
\begin{align}
    \braket{X_{m,-m}(\tau)} = e^{-4m^2 f_z(\tau)},
\end{align}
where
\begin{align}
    f_z(\tau) &= \frac{\tau^2}{4\pi}\int_{-\infty}^{\infty}S_z(\omega)\text{sinc}^2(\omega \tau/2) d\omega. \label{eq:fz}
\end{align}
On the other hand, for an initially-prepared superposition state
\begin{align}
    \ket{\psi(0)} = \frac{1}{\sqrt{2}}\left(\ket{m}+\ket{m+1}\right),
\end{align}
The coherence is given by
\begin{align}
     \braket{X_{m,m+1}(\tau)}=e^{-f_z(\tau) - (2m+1)^2f_{zz}(\tau)},
\end{align}
where $f_z(\tau)$ is defined as in Equation \ref{eq:fz}, and
\begin{align}
    f_{zz}(\tau) &= \frac{\tau^2}{4\pi}\int_{-\infty}^{\infty}S_{zz}(\omega)\text{sinc}^2(\omega \tau/2) d\omega.
\end{align}

\begin{figure}[!h]
    \centering
    \includegraphics[width=0.65\linewidth]{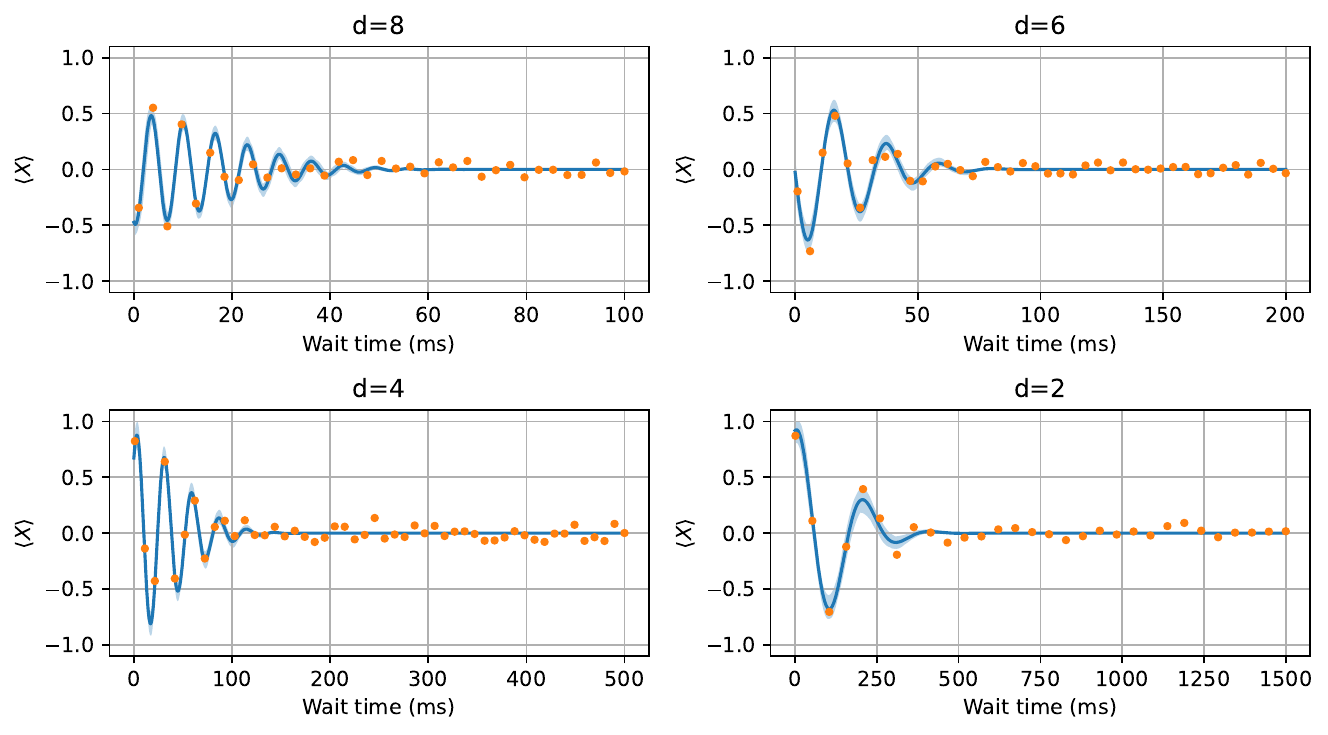}
    \caption{\textbf{The fitting of the cat-subspace Ramsey measurements for Device A}. The shading represents the distribution of the fit over the 2000 runs of bootstrapping. }
    \label{fig:cat_A}
\end{figure}

\begin{figure}[!h]
    \centering
    \includegraphics[width=0.95\linewidth]{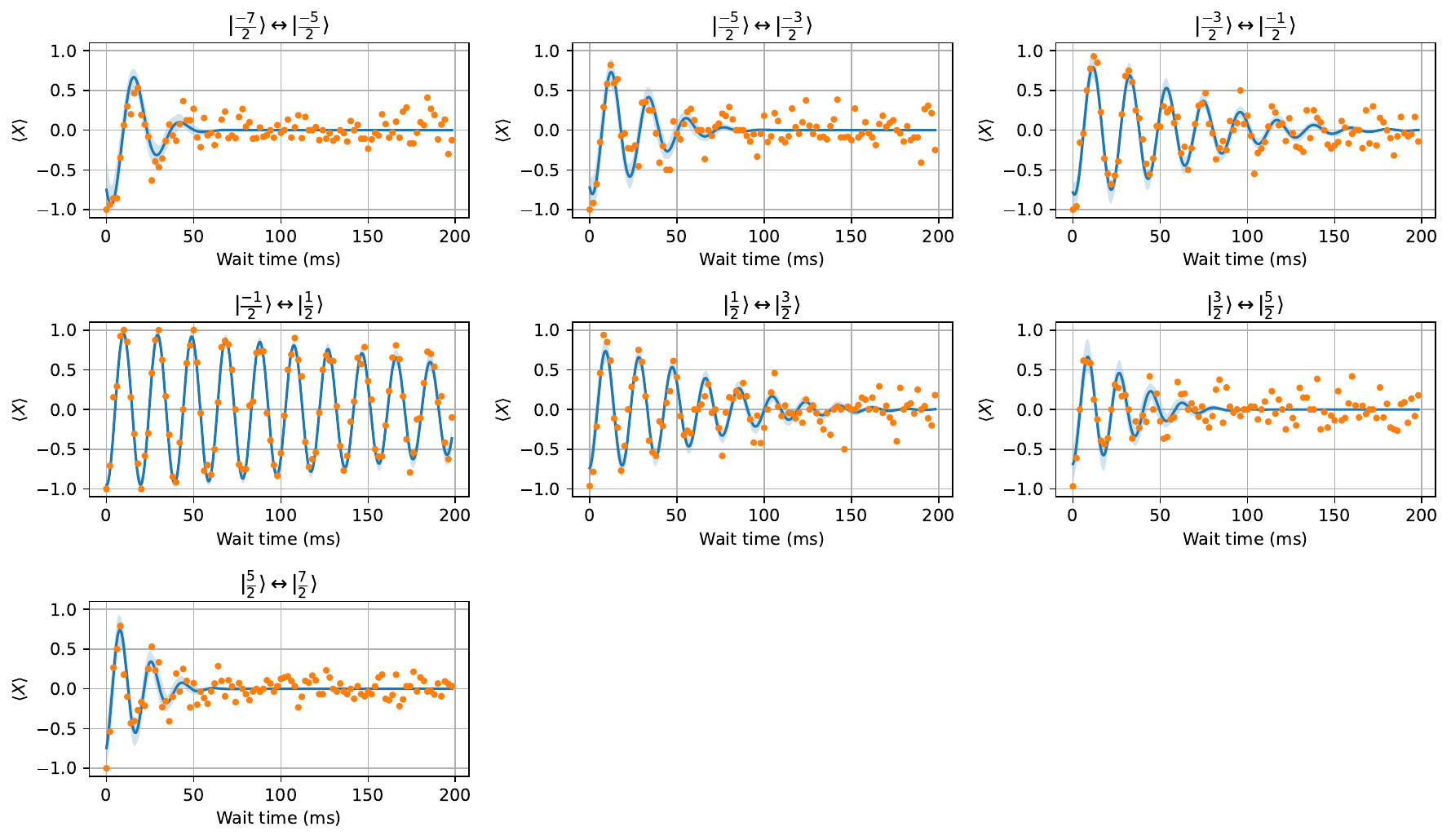}
    \caption{\textbf{The fitting of the adjacent-subspace Ramsey measurements for Device A}. The shading represents the distribution of the fit over the 2000 runs of bootstrapping. }
    \label{fig:subspace_A}
\end{figure}

Given the experimental Ramsey measurements, we can fit a model for the PSD of the noise processes. We choose them to be in the form 
\begin{align}
    S_z(\omega) &= \frac{A_M}{\omega^{\alpha_M} } \\
    S_{zz}(\omega) &= \frac{A_E}{\omega^{\alpha_E} }
\end{align}
To avoid the singularity at DC frequencies, we choose a low cutoff frequency $\omega_c$, effectively rendering the PSD zero for $|\omega|<\omega_c $. To account for SPAM errors, we use the modified models:
\begin{align}
    \braket{\tilde{X}_{m,-m}(\tau)} = a_{m,-m} e^{-4m^2 f_z(\tau)}\cos(\Omega_{m.-m} \tau + \phi_{m,-m}) \\
    \braket{\tilde{X}_{m,m+1}(\tau)}=  a_{m,m+1} e^{-f_z(\tau) - (2m+1)^2f_{zz}(\tau)} \cos(\Omega_{m,m+1} \tau + \phi_{m,m+1}),
\end{align}
The parameters $A_{M}, A_{E}, \Omega_{m,-m},$ and $\Omega_{m,m+1}$  are constrained to be non-negative, the parameters $a_{m,-m}, a_{m,m+1}$ are constrained to be in the interval $[-1,1]$, and the parameters $\alpha_{M}, \alpha_{E}$ are constrained to be greater than 1. Here, we do a two-stage fitting procedure. First, we fit the cat-subspace data, since the evolution is only governed by magnetic noise. This gives us information about $S_z(\omega)$. We do that via a joint fit between the four cat states $m=\frac{7}{2},\frac{5}{2}, \frac{3}{2}, \frac{1}{2}$. In this joint fit, we assume that $A_1$ and $\alpha_1$ are exactly the same across the four states, but all the other parameters vary from one state to another. This results in a model with total trainable parameters of 14 (4 scaling amplitudes $a_{m,-m}$, 4 rotations frequencies $\Omega_{m,-m}$, 4 phase shifts $\phi_{m,-m}$ all accounting for spam, in addition to 2 parameters for the noise spectrum). We implement the model using Tensorflow \cite{tensorflow} and Keras \cite{keras} Python libraires, and use ADAM optimization \cite{Adam} for the fitting, with the cost function defined to be the mean square error (MSE) between the model prediction and actual data, averaged over the 4 states. In the second stage of the fitting, we focus on the adjacent subspace data. Here we take the value of $\alpha_M$ obtained from the first fit, and fix it in the adjacent subspace model, effectively rendering it non-trainable. This reflects our intuition that $\alpha_M$ being a fundamental property of the noise should not change significantly over the course of running the experiment, since the device and its environment are the same. However, we allow the $A_M$ to change as the strength of the noise could drift over time. Thus, by isolating the effect of the magnetic noise using the cat-subspaces, this fit can give us more information about the electric noise.  Since we have 7 states to prepare, the total number of trainable parameters is 24 (7 scaling amplitudes $a_{m,m+1}$, 7 rotations frequencies $\Omega_{m,m+1}$, 7 phase shifts $\phi_{m,m+1}$ all accounting for SPAM, in addition to 3 parameters for the noise spectra). Once again we use ADAM optimizer, and MSE over the seven states as cost function. After fitting, we can isolate the effect of SPAM from the noise spectra. Thus, we can find $T_2^*$ simply by using bisection search algorithm, based on the fact that $f_z(0)=0, f_{zz}(0)=0$, and the definition of $T_2^*$,
\begin{align}
    \braket{X_{m,-m}(T_2^*)}&:=e^{-1} \implies 4m^2f_z(T_2^*)=1, \\
\end{align}
for cat-subspace and 
\begin{align}
        \braket{X_{m,m+1}(T_2^*)}&:=e^{-1} \implies f_z(T_2^*)+(2m+1)^2f_{zz}(T_2^*)=1. \\
\end{align}
for adjacent subspace. Finally, to get the error bars for the fits, we utilize a bootstrapping approach as follows:
\begin{enumerate}
    \item Fit the model using the aforementioned procedure.
    \item Compute the error residuals at each point, i.e. the difference between the model prediction and the actual data point
    \item Create 2000 artificial datasets, each constructed by adding a random permutation of the sequence of error residuals to the model predictions to generate an artificial sequence of data points.
    \item Refit the model using the optimal parameters from step 1 as initial guess for each of the 2000 datasets.
    \item Calculate $T_2^*$ for each of the 2000 fits, as well as the model predictions.
    \item Compute required statistics from the results of step 5. 
\end{enumerate}
In obtaining the error bars for the subspace fit, we fix $\alpha_M$ using the optimal fit rather than using the values we get from each bootstrapping run over the cat-subspace. In other words, we treat the two fitting as separate fits when finding the error bars, in which case the second fit will have one fixed non-trainable $\alpha_M$. To get error bars of $T_2^*$, we recompute its value for each of the 2000 fits we get from bootstrapping. The resulting noise spectrum fitting parameters for Devices A and B are shown in Table \ref{tab:fit_parameters}. Figures \ref{fig:cat_A} and \ref{fig:subspace_A} show the obtained fits for Device A, as well as the full distribution of the 2000 fits obtained by bootstrapping. Figures \ref{fig:cat_B} and \ref{fig:subspace_B} show similar plots for Device B.

\begin{table}
    \centering
    \renewcommand{\arraystretch}{1.25}
   \begin{tabular}{|l|l|ll|ll|}
    \hline
                                   & Parameter & \multicolumn{2}{c|}{Device A} & \multicolumn{2}{c|}{Device B} \\ \hline
    \multirow{2}{*}{Cat-subspace}      & $\alpha_M$  &  1.0   &  [1.0, 1.34]  &   1.0    & [1.0 1.43]              \\ \cline{2-6} 
                                       & $A_M$       &  26.69 & [5.12, 32.99] &   29.01  & [5.75, 31.80]              \\ \hline
    \multirow{3}{*}{Adjacent-subspace} & $\alpha_E$  &  1.0   & [1.0, 1.51]   &   1.0    & [1.0, 1.42]              \\ \cline{2-6} 
                                       & $A_M$       &  21.84 & [14.17,29.13] &   59.68  & [38.59, 74.57]              \\ \cline{2-6} 
                                       & $A_E$       &  46.0  & [8.25, 51.74] &   38.08  & [12.11, 41.48]              \\ \hline
    \end{tabular}
    \caption{\textbf{The fitting parameters.} The numbers between the brackets represent the 95\% confidence interval. }
    \label{tab:fit_parameters}
\end{table}

\begin{figure}[!h]
    \centering
    \includegraphics[width=\linewidth]{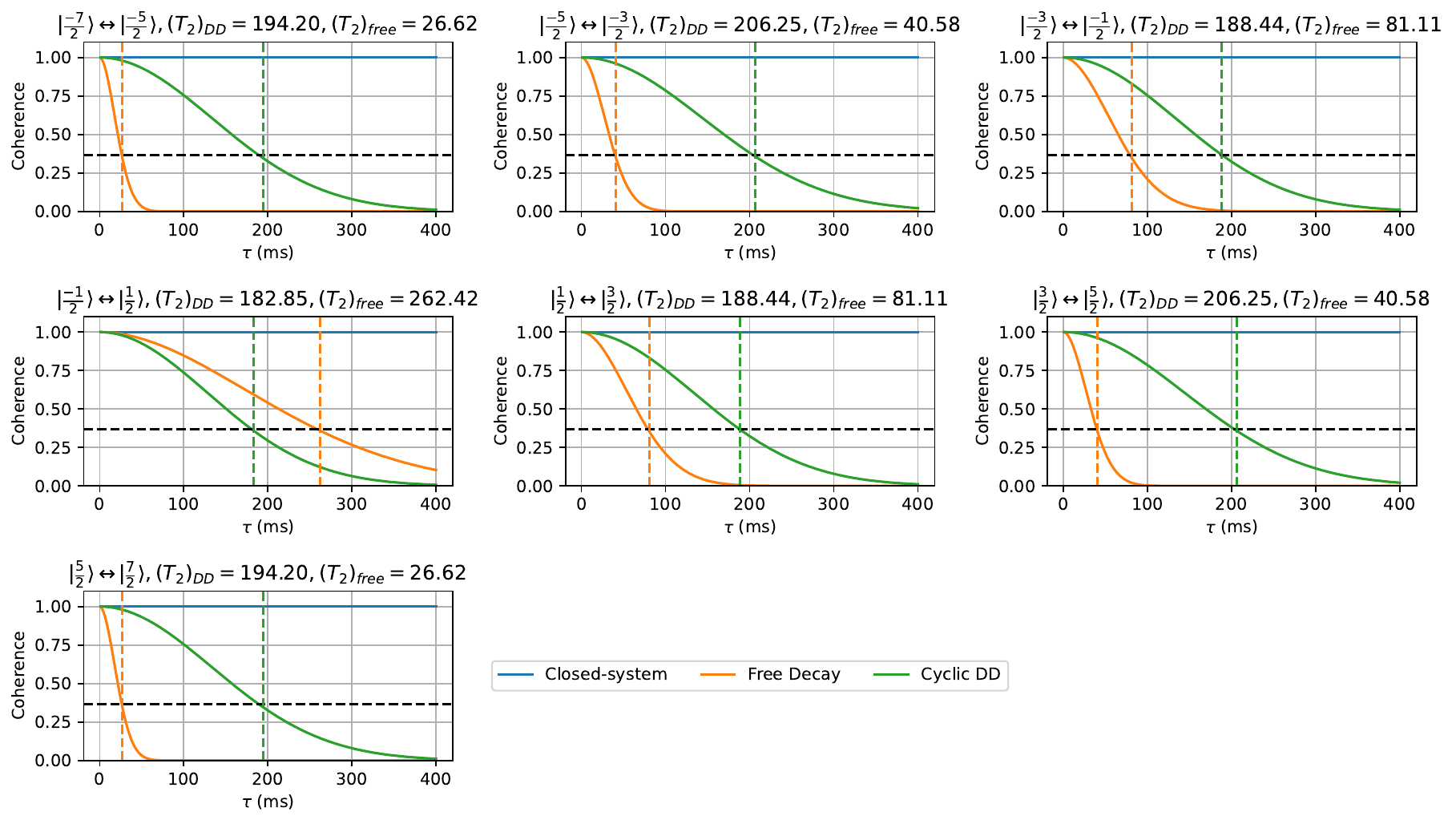}
    \caption{\textbf{Simulation of the free-decay vs cyclic DD for the extracted noise PSD for Device A.}}
    \label{fig:subspace_sim}
\end{figure}

\begin{figure}[!h]
    \centering
    \includegraphics[width=0.65\linewidth]{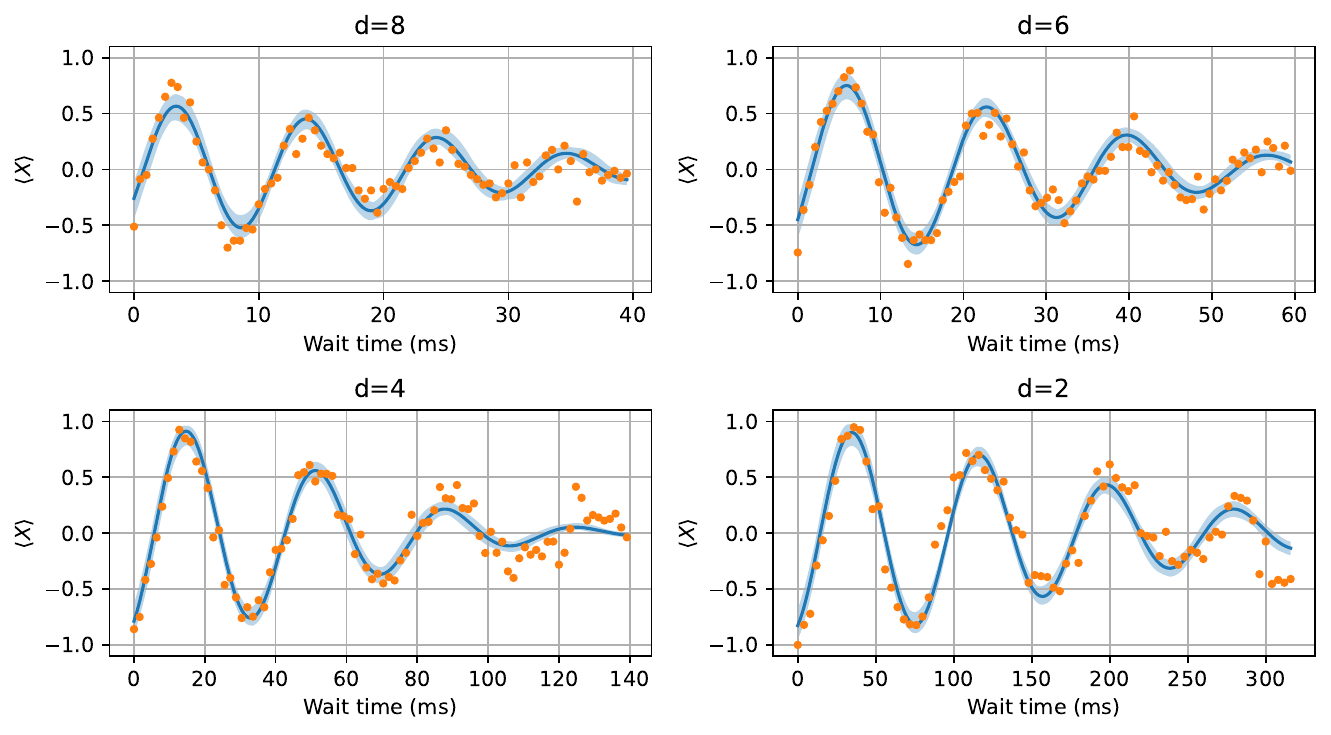}
    \caption{\textbf{The fitting of the cat-subspace Ramsey measurements for Device B}. The shading represents the distribution of the fit over the 2000 runs of bootstrapping. }
    \label{fig:cat_B}
\end{figure}

\begin{figure}[!h]
    \centering
    \includegraphics[width=0.95\linewidth]{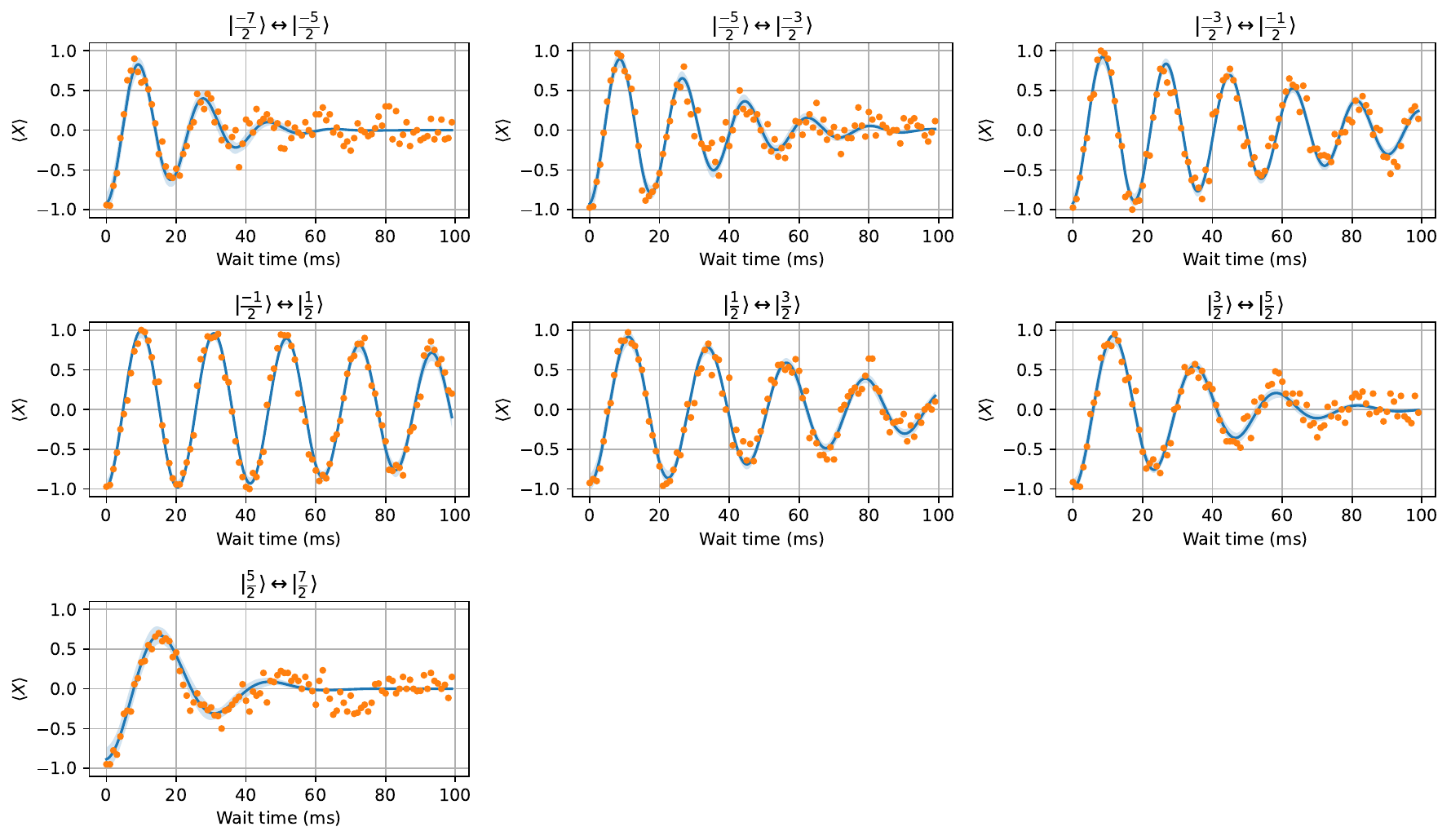}
    \caption{\textbf{The fitting of the adjacent-subspace Ramsey measurements for Device B}. The shading represents the distribution of the fit over the 2000 runs of bootstrapping. }
    \label{fig:subspace_B}
\end{figure}

After obtaining the fit parameters for the adjacent-subspace states, we can then utilize the fitted PSD to simulate the effect of cyclic DD sequence. Here, we assume an ideal control sequence (i.e. Dirac Delta impulses). Together with the assumptions on the noise process, one can obtain an exact expression using filter functions \cite{theory_paper}. We use those expressions to plot obtain the value of $T_2^*$ under the effect of cyclic DD and compare it with the values for free decay. The error bars for the cyclic DD $T_2^*$ are obtained by re-running the simulations for each of the 2000 fits we get from the bootstrapping, and then find the $T_2^*$ numerically for each run of the simulation. Figure \ref{fig:subspace_sim} shows the result of the simulations. Finally, this method in general can yield different suboptimal models for the PSD of the noise depending on the initialization of the fitting parameters and constraints. Further studies based on model selection techniques \cite{model_selection} or performing standard quantum noise spectroscopy \cite{QNS} could provide a global optimal solution to the fitting.  


\FloatBarrier

\section{DD strategies with static noise}

We illustrate the selective cancellation of different noise terms with the four separate DD strategies employed in Fig.~5. We compare the 4 DD methods under the influence of two types of static noise: $I_z$ and $I_z^2$. We simulate the propagator over a total time $T=\SI{70}{\milli \second}$, as a function of the static noise coupling strength (0-10~Hz). For the continuous strategies, spin locking, and rotary echo, we simulate a drive strength of $\Omega_R=2\pi \times \SI{100}{\hertz}$. The fixed-length pulses are for simplicity assumed to be instantaneous, therefore, this simulation is to be interpreted as illustrative rather than quantitatively accurate.

\begin{figure}[h]
    \centering
    \includegraphics[width=\linewidth]{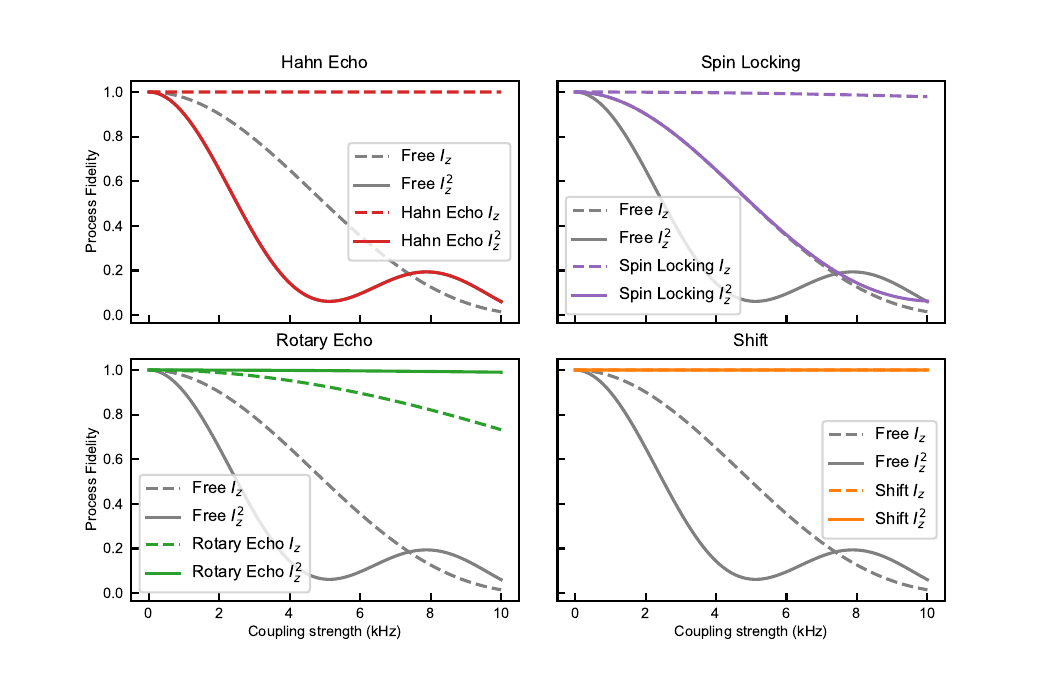}
    \caption{\textbf{Process fidelity with respect to the identity operator under the influence of static $I_z$ (dashed) or $I_z^2$ noise (solid lines).} We compare the four DD strategies employed in Fig.~5 of the main text. Grey lines show the bare, unprotected fidelity for reference. 
    \textbf{a)} The SU(2) Hahn echo cancels static $I_z$ noise.
    \textbf{b)} The rotary echo applies a continuous global $I_x$ drive sandwiched in between two $\pi/2$ $I_y$ rotations. It primarily cancels $I_z^2$ noise, but also protects against $I_z$ noise.
    \textbf{c)} Spin locking is particularly efficient against $I_z$ noise, and provides protection to a lesser extent against $I_z^2$. 
    \textbf{d)} In the ideal pulse limit, the shift-DD sequence cancels both static $I_z$ and $I_z^2$. 
    }
    \label{fig:placeholder}
\end{figure}


\section{Optimization of Spin Locking power}

\begin{figure}[h]
    \centering
    \includegraphics[width=\linewidth]{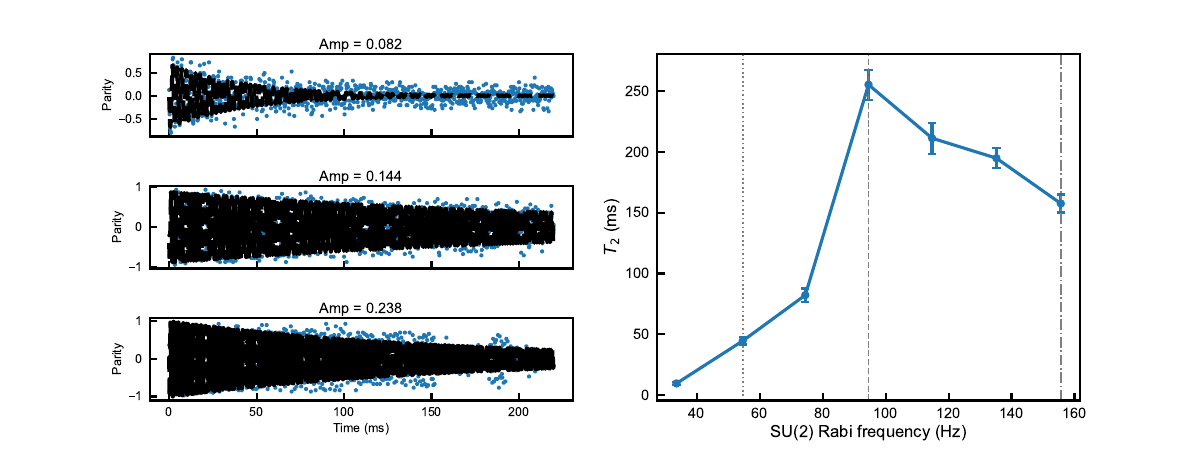}
    \caption{\textbf{Optimizing the spin locking drive power}. The measurement data shows parity oscillations vs. amplitude for three selected amplitudes. The decay constant $T_2$ of the parity oscillations is maximized for Amp=0.144, corresponding to an SU(2) Rabi frequency of \SI{95}{\hertz}.}
    \label{fig:parity_vs_amp}
\end{figure}

We optimize the power of the spin locking drive by measuring parity oscillations of the spin-cat state $\ket{+}_L$ oriented along the $I_x-$axis. By continuously rotating with an $I_x-$drive, the parity oscillates between 1 and -1 before decoherence reduces the parity oscillation amplitude. In Fig.~5 of the main text we chose an amplitude setting of Amp=0.15, close to the optimal amplitude setting (Amp=0.144, $\Omega_R=\SI{95}{\hertz}$, see Fig.~\ref{fig:parity_vs_amp}).

\FloatBarrier
\bibliography{bib}